\documentclass{aa}  
\usepackage{natbib}
\usepackage[font=small,labelfont=bf]{caption}
\bibpunct{(}{)}{;}{a}{}{,} 
\newcommand{\rmd}{\mathrm{d}}
\usepackage{graphicx}
\usepackage{color}
\usepackage{txfonts}
\begin{document}

   \title{Observational signatures of magnetic field structure in relativistic AGN jets}
\titlerunning{Observational signatures of magnetic field structure in relativistic AGN jets}

   \author{Christopher Prior
          \inst{1}
          \and
          Konstantinos N. Gourgouliatos\inst{2}
          }

   \institute{Department of Mathematical Sciences, Durham University\\
              \email{christopher.prior@durham.ac.uk}
         \and
             Department of Mathematical Sciences, Durham University\\
              \email{konstantinos.gourgouliatos@durham.ac.uk}
             }
\authorrunning{C. Prior \& K.N. Gourgouliatos}
   \date{Received ---; accepted ---}

 
  \abstract
   {Active Galactic Nuclei (AGN) launch highly energetic jets sometimes outshining their host galaxy. These jets are collimated outflows that have been accelerated near a  supermassive black hole located at the centre of the galaxy. Their, virtually indispensable, energy reservoir is either due to gravitational energy released from accretion or due to the extraction of kinetic energy from the rotating supermassive black hole itself. In order to channel part of this energy to the jet, though, the presence of magnetic fields is necessary. The extent to which these magnetic fields survive in the jet further from the launching region is under debate. Nevertheless,  observations of polarised emission and Faraday rotation measure confirm the existence of large scale magnetic fields in jets.}
   {Various models describing the origin of the magnetic fields in AGN jets lead to different predictions about the large scale structure of the magnetic field. In this paper we study the observational signatures of different magnetic field configurations that may exist in AGN jets in order to asses what kind of information regarding the  field structure can be obtained from radio emission, and what would be missed. }
   {We explore three families of magnetic field configurations. First, a force-free helical magnetic field corresponding to a dynamically relaxed field in the rest frame of the jet. Second, a magnetic field with a co-axial cable structure arising from the Biermann-battery effect at the accretion disk. Third, a braided magnetic field that could be generated by turbulent motion at the accretion disk. We evaluate the intensity of synchrotron emission, the intrinsic polarization profile and the Faraday rotation measure arising from these fields. We assume that the jet  consists of a relativistic spine where the radiation originates from and a sheath containing thermalised electrons responsible for the Faraday screening. We evaluate these values for a range of viewing angles and  Lorentz factors. We account for Gaussian beaming that smooths the observed profile. }
  {Radio emission distributions from the jets with dominant large-scale helical fields show asymmetry across their width. The Faraday rotation asymmetry is the same for field's with opposing chirality (handedness). For jets which are tilted towards the observer the synchrotron emission and fractional polarization can distinguish the fields chirality. When viewed either side-on or at a Blazar type angle only the fractional polarization can make this distinction. Further this distinction can only be made if the direction of the jet propagation velocity is known, along with the location of the jet's origin. The complex structure of the braided field is found not to be observable due to a combination of line of sight integration and limited resolution of observation. This raises the possibility that, even if asymmetric radio emission signatures are present, the true structure of the field may still be obscure.}
   {}

   \keywords{giant planet formation --
                $\kappa$-mechanism --
                stability of gas spheres
               }

   \maketitle
%

\section{Introduction}

AGN are the most luminous long-lived sources in the Universe, outshining their host galaxies and persisting for millions of years \citep{Fabian:1999}. Their power can exceed $10^{46}$~erg~s$^{-1}$ and originates from a central engine consisting of a supermassive black hole and an accretion disc \citep{Salpeter:1964, Zeldovich:1964}. While the gravitational energy released from the material accreted to the black hole is sufficient to match the energetics of the jet, the actual formation of the jet is a highly non-trivial process. Pure hydrodynamical models cannot explain  jet launching. The role of a magnetic field in the vicinity of the black hole has been stressed from the very early works about these sources \citep{Lynden-Bell:1969}. The formation of the jet could be related to a dynamo action at the accretion disk which leads to the formation of two oppositely propagating magnetic beams \citep{Lovelace:1976}. 

The magnetic field has also a central role in the Blandford-Znadjek mechanism that describes the extraction of energy from a Kerr black hole through a magnetic field that powers a jet \citep{Blandford:1977}. Alternatively if the magnetic field is coupled to the accretion disc the energy can be extracted through the Blandford-Payne mechanism \citep{Blandford:1982}. These  mechanisms do not depend on the polarity of the magnetic field with respect to the flow direction. If, on the other hand, the magnetic field originates from a cosmic battery phenomenon \citep{Contopoulos:1998}, then the polarity of the magnetic field will depend on the jet flow direction. In this model the electrons of the accretion disk experience a Poynting- Robertson drag, slowing them down with respect to the disk flow. This generates a net current and flux loops. Due to differential rotation, these loops open up, while accretion accumulates flux of one polarity at the inner part of the disk and the opposite polarity at the outer edge. This leads to the eventual formation of a toroidal magnetic field that corresponds to an electric current anti-parallel to the jet flow in the inner part of the jet, and parallel to the flow in the outer part \citep{Contopoulos:2006}. The overall picture resembles that of a cosmic co-axial cable with electric current flowing towards the origin of the jet at the inner part of the cable and in the opposite direction in the outer part of the cable \citep{gabuzda2018jets}.

A magnetic field anchored to an accretion disk that rotates rapidly might be twisted in similar manner, thus one might expect it to generate helical magnetic field \citep{Spruit:2010}. One assumption made in modelling this possibility is that the jet can be described by a close to ideal magnetohydrodynamical fluid with a magnetic field whose Lorentz force is the dominant driver of the field's evolution \textit{e.g.} \cite{konigl1985force}. Following an argument of  \cite{taylor1974relaxation}, magnetic re-connection at sites where entanglement of the field lines is complex, and hence local the current is high, means that only large scale twisting induced by the rotation of the field will survive the field's relaxation (the average magnetic helicity of the field is conserved). Under this assumption the system should relax to a force-free equilibrium, which, in the fluid frame where the electric field vanishes, takes the form
\begin{equation}
\label{forcefreeq}
\nabla \times {\bf B} =\alpha {\bf B}
\end{equation}
with $\alpha$ a constant in space. This argument was applied to Jets by \cite{konigl1985force}. Solutions of (\ref{forcefreeq}) in cylindrical geometry lead to configurations such as the reverse field pinch \citep{Lundquist:1951} were used to interpret emission data in \cite{clausen2011signatures}. 

Since the possibility of turbulent motion in the accretion disc has long been considered viable  \citep{Stella:1984, Subramanian:1996,balbus1998instability,carballido2005diffusion,lesur2005relevance} it seems sensible to consider the possibility that this motion could impart itself into the jet's magnetic field leading to more complex magnetic field configurations than a helical field. In such a system magnetic field lines anchored onto the accretion disk might experience extra local twist in addition to the global differential rotation. Even if the force free condition holds significantly complex structures can arise in Eq.~\ref{forcefreeq} if $\alpha$ is allowed to vary in space (this has long been considered in Solar coronal modelling which relies on similar {\bf assumptions \textit{e.g.}} \citealt{derosa2009critical,su2009observations,wiegelmann2012solar}). Moreover, the inclusion of the plasma thermal pressure leads to further modifications of the force-free equilibrium \citep{Gourgouliatos:2012}. Attempting to simulate this possibility would be a complex task so we consider here a sensible first step to see if the possibility of significantly complex magnetic fields can be reconciled with observational data. 

While the magnetic field is critical for the dynamics of the jet it is also the catalyst for the generation of the observed non-thermal emission. This comes through two main effects: synchrotron radio emission and Faraday rotation. Synchrotron radiation is produced when charged particles gyrate round magnetic field lines and experience acceleration. Because of this acceleration, these particles generate polarised radio waves \citep{Rybicki:1986}. The intensity and polarisation of the synchrotron emission received by an observer depends on the population of relativistic electrons and the intensity of the magnetic field perpendicular to the line-of-sight vector, the vector originating from the observer to the source of radiation. Once polarised emission crosses a thermalised optically thin electron gas containing a magnetic field, the polarisation plane shifts due to Faraday rotation. The level of Faraday rotation is proportional to the integral of the magnetic field component along the line-of-sight, the thermal electron number density and the square of the wavelength of the radio emission. Thus, by taking observations at different wavelengths, one can measure the value of the rotation measure.

Therefore, the effects of the jet magnetic field can be indirectly probed through radio observations. This information depends on the line-of-sight integrated effect of the magnetic field and obviously information is lost due to the two dimensional projection of a three dimensional structure. In addition to the geometric effects, the highly relativistic nature of astrophysical jets, makes interpretation more complicated due to relativistic aberration and Doppler shifting. Because of these effects, the angle between the line-of-sight vector and the jet velocity changes and needs to be accounted for when observable properties are to be extracted. Finally there is the matter of the limited resolution due to the finite beam size of the radio telescope \citep{Hovatta:2012}. This leads to the loss of the various emission features due to convolution of the actual signal with the radio beam \citep{clausen2011signatures}.

Extensive imaging surveys using very long base interferometry has permitted milliarcsecond resolution \citep{Zensus:1997} radio-imaging of the central part of the AGN to sub-parsec scales {\it e.g.}~\citep{Walker:2018}. Such observations have revealed transverse asymmetries in the polarisation and Faraday rotation measure implying that a large scale ordered magnetic field is present in the jet \citep{asada2002helical,Mahmud:2013, Gabuzda:2015, Motter:2017}. While the presence of the magnetic field in jets is indisputable, its exact structure is yet to be resolved \citep{Hovatta:2018}. 

In this paper we explore three main families of magnetic field configuration, extracting their observational properties. Two are based on fields that have previously been proposed to exist in relativistic jets, the reverse pinch field  ascribed to the Taylor relaxation hypothesis \citep{konigl1985force,clausen2011signatures} and a representation of coaxial field of the type proposed in \cite{contopoulos:2009,gabuzda2018jets}. There is no analytical expression for the magnetic field for the battery/coaxial cable model, so we have proposed simple representation in order to compare its expected observations to those of the previously studied reverse-field pinch. The aim is to see if they could be distinguished. The third field is a braided field used in \cite{pontin2011dynamics,wilmot2010dynamics,prior2016twisted} as the initial configuration of a field which relaxes to a non-Taylor state (as discussed above). This last field is use in this note not as a realistic proposed model of the jet's field, but as an indication of the variety structures which could explain radio observations. What is crucial here is that the size and strength of the braided field structure is such that it would significantly affect the evolution of a large scale magnetic field. This field, whilst complex is also significantly organised and large scale, thus differs from the class of disorganized fields used in \cite{laing1981magnetic}.

In section 2 we present a geometrical model of the observation of the jet and the specific models used for the jet's magnetic field. We also discuss the appropriate relativistic transformations to produce synthetic images of the jets emission properties. In section 3 we present the results of a survey of synthetic observational signals of the various magnetic field configurations, viewing angles and jet velocities. In section 4 we consider some mathematical aspects of the loss of information due to the finite beam width which give some insight as to the type of emission structures which cannot be detected using this method. In section 5 we compare our results to some observations and in section 6  we conclude.

\section{The toy jet radio emission model}
\subsection{Basic geometry}
\label{BASIC_GEOMETRY}
\begin{figure}
    \centering
   (a)\includegraphics[width=8cm]{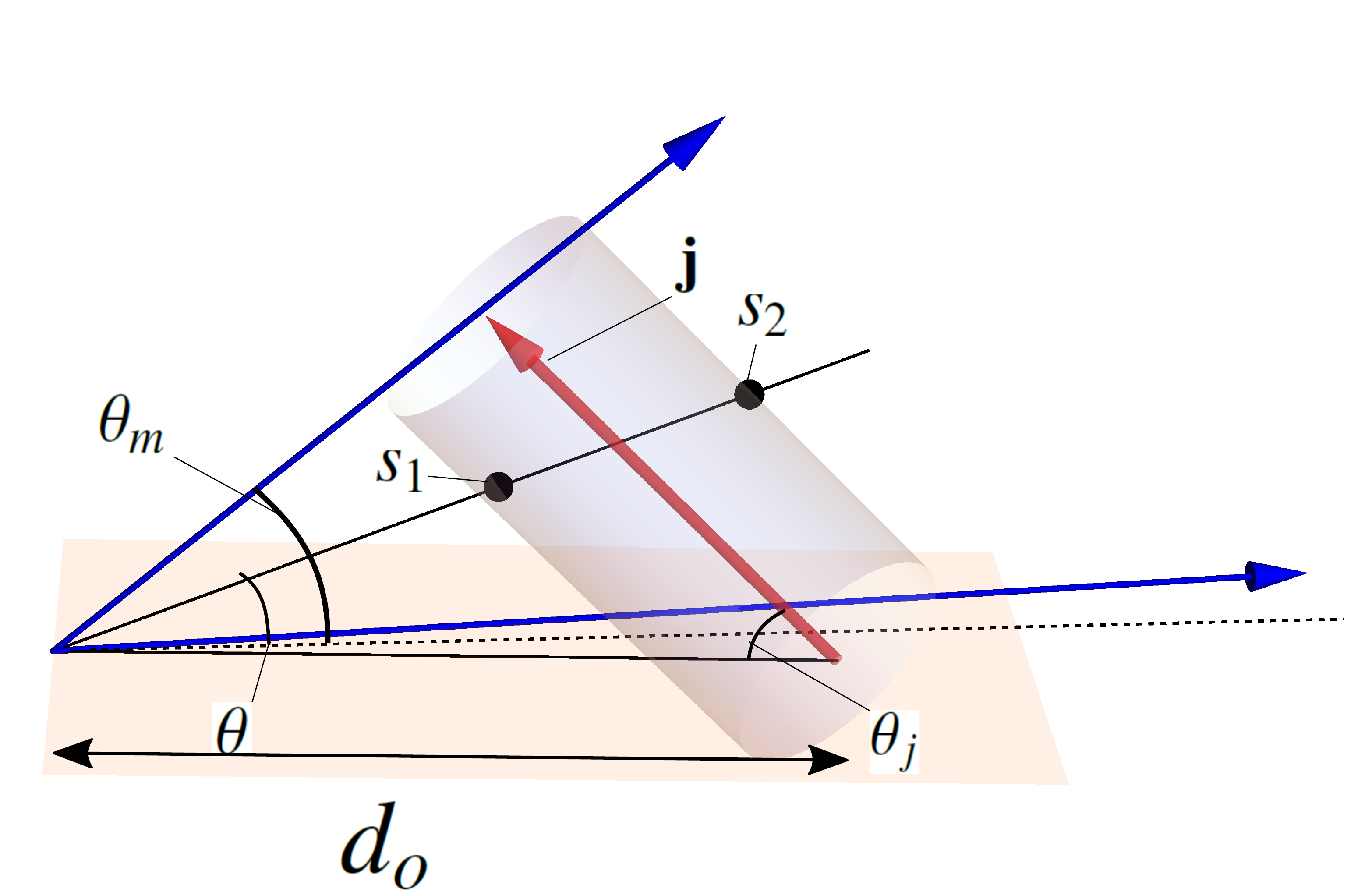}\quad (b)\includegraphics[width=9cm]{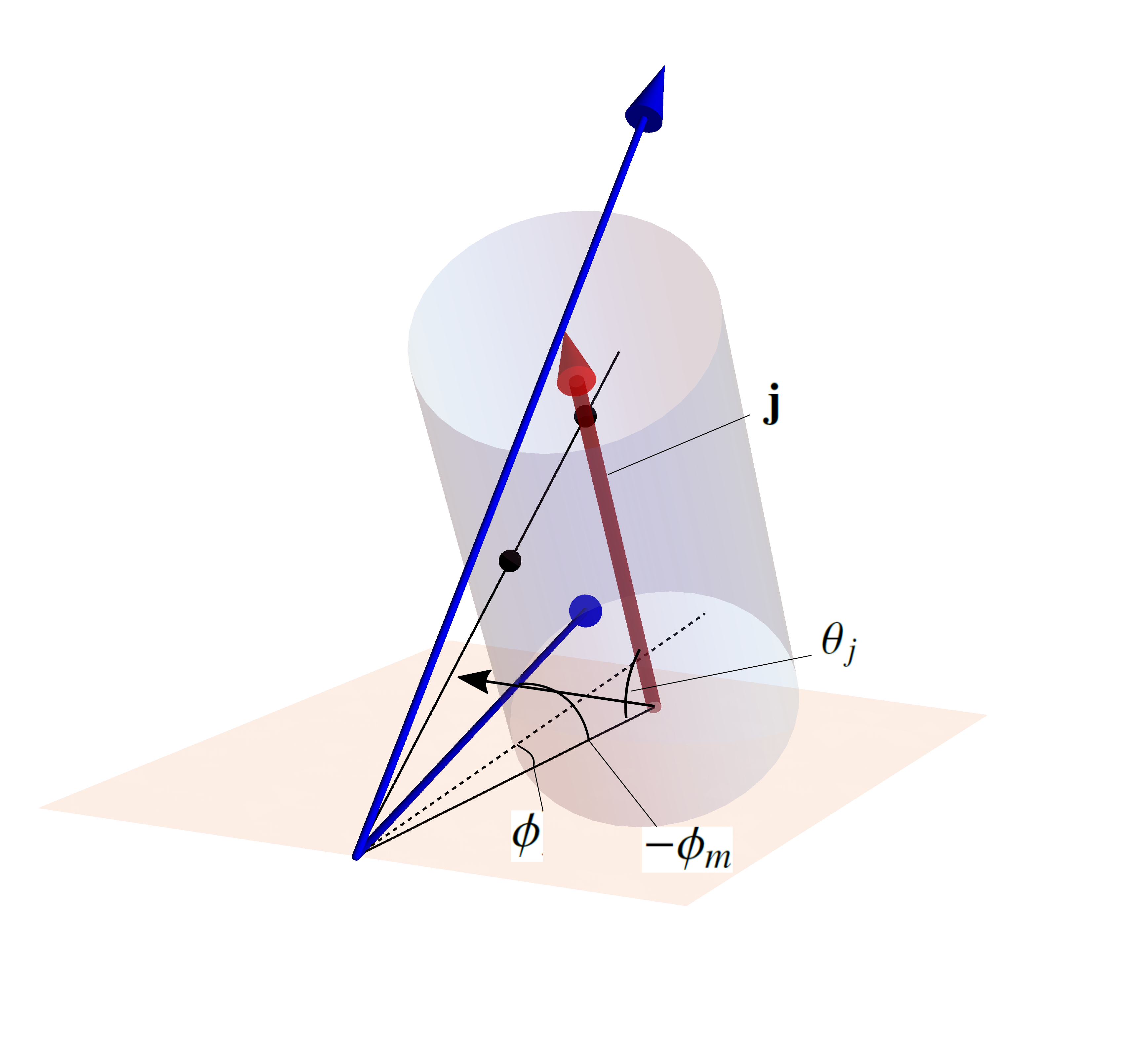}
    \caption{Representations of the model geometry. In (a) we see the model as viewed along the $y$-axis. In (b) the same figure is seen from "over the shoulder" of the observer. The jet direction ${\bf J}$ is shown as a red arrow, the jet angle $\theta_j$ this makes with the $x$-axis is also shown. The distance $d_0$ from the observer to the jet along the $x$ axis is shown. The Blue arrow skimming the jet's upper surface indicates the maximum angle subtended by the jet to the viewer $\theta_m$. A black line, representing the direction of photon observation, pierces the cylinder surface at $s_1$ and $s_2$ subtends an angle $\theta<\theta_m$. The second blue arrow indicates the maximum (in magnitude) $\phi$ angle $-\phi_m$ subtended by the viewer and the jet's side edge, as indicated in (b). Also shown in (b) as a dotted line is the projection of the photon observation direction into the $x$-$y$ plane and the angle $\phi$ which indicates its off-set from the jet's central axis.}
    \label{modelgeom}
\end{figure}
The model geometry is indicated in Figure \ref{modelgeom}.
We consider one arm of a jet whose origin coincides with the origin of a Cartesian coordinate system $(x,y,z)$. An observer is placed at $(-d_o,0,0)$ and the axis of the jet, the direction of its bulk velocity (magnitude $\beta c$ with $c$ the speed of light) ${\bf v}$ lies in the $y$-$z$ plane ${\bf v} = \beta c\left(-\cos{\theta_j},0,\sin\theta_j\right)$, with $\theta_j=0$ meaning the jet is pointing towards the observer and $\theta_j=\pi$ pointing away. We assume the field is defined within a cylinder of radius $1$ and height $h$. 
\subsection{Viewing angles}
The direction of observation of the field is determined by the intersection of a set of lines $l(s,\theta,\phi)$, with
\begin{equation}
l(s,\theta,\phi) =(-d_o,0,0) + s\left(\cos\theta\cos\phi,\cos\theta\sin\phi,\sin\theta\right).
\end{equation}
and consequently the direction of travel of the photons observed at the angle pair $(\theta,\phi)$ is 
\begin{equation}
{\bf n} = -\left(\cos\theta\cos\phi,\cos\theta\sin\phi,\sin\theta\right).
\end{equation}
For each pair $(\theta,\phi)$ the line $l(s,\theta,\phi)$ will either intersect the jet cylinder twice or never, except where it skims the boundary. If the solution exists one obtains a pair $s_1(\theta,\phi)$ and $s_2(\theta,\phi)$ where the line enters and leaves the jet. Most of the emission quantities we calculate will be integrals over the line $\left(l(s,\theta,\phi)\vert s\in[s_1(\theta,\phi),s_2(\theta,\phi)]\right)$. Thus we have a viewing domain ${\cal D} = \left\{(\theta,\phi)\vert\, \theta\in[0,\theta_m],\phi\in[-\phi_m,\phi_m]\right\}$, with $\theta_m$ and $\phi_m$ the maximum viewing angles across the jet's vertical extent and its width. When presenting the results we map this range of angles to the projected distances $\phi_d =\sin{\phi}/\sin \phi_{max}$ and $\theta_d =\sin\theta/\sin\theta_{max}$ with the normalisation for clarity \textit{i.e.} $\theta_d=0.5$ is the viewing angle at half of the maximum vertical viewing extent. 

For the calculations in this paper the ratio $d_o/h$ will typically be large so the actual range of viewing angles $\arccos({\bf n}(\theta,\phi)\cdot {\bf v})$ does not differ too much from those of the value of $\theta_j$ (allowing for the fact that they will be transformed due to relativistic aberration).

We now develop the models for the various radio emissions we simulate. The following largely follows the models described in \cite{lyutikov2003polarization,lyutikov2005polarization,clausen2011signatures}.
\subsection{Stokes parameters and linear polarization}
We calculate the following Stokes parameters
\begin{align}
I &= C_1\int_{s_1}^{s_2}D^{2+(p+1)/2}\vert {\bf B}'\times{\bf n}'\vert^{\frac{p+1}{2}}\rmd{s},\\
\nonumber Q&= C_2 \int_{s_1}^{s_2}D^{2+(p+1)/2}\vert {\bf B}'\times{\bf n}'\vert^{\frac{p+1}{2}}\cos(2\zeta)\rmd{s},\\
\nonumber U&= C_2 \int_{s_1}^{s_2}D^{2+(p+1)/2}\vert {\bf B}'\times{\bf n}'\vert^{\frac{p+1}{2}}\sin(2\zeta)\rmd{s},\\
\nonumber V&=
0.
\end{align}
where a {\bf prime} indicates the quantity is expressed in the jet rest frame, here the integration is in the observer frame. The constant $p$ is the electron index. The electron index depends on the process that has accelerated the electron population \citep{Pacholczyk:1970, Drury:1983}, we choose a value $1.5$ in this text, changing this value within a reasonable range ($1.5<p<3$) does not significantly affect the results.
All integrals are along the line of sight ${\bf l}(s,\theta,\phi)$. Since we are only interested in the qualitative comparison of distributions of these quantities the precise values of $C_1,C_2$ are not critical. The fields ${\bf B}$ will generally be specified in the jet frame (${\bf B}'$). The jet frame observation direction ${\bf n}'$ is the aberration corrected (unit vector) direction along which the field is viewed, with
\begin{align}
{\bf n}' =\frac{(1+\gamma){\bf n} + \gamma^2({\bf n}\cdot {\bf v}){\bf v}}{(1+\gamma)\sqrt{1+ \gamma^2({\bf n}\cdot {\bf v})^2}}, 
\end{align}
(see \textit{e.g.} \cite{lyutikov2003polarization}). The function $D =1/\gamma(1-{\bf v}\cdot{\bf n})$ is the Doppler boosting factor, where $\gamma$ is the Lorentz factor $(1-\beta^2)^{1/2}$. We introduce unit vector ${\bf l}$ normal to the plane containing the observation direction ${\bf n}$ and the reference  direction in the plane of the sky, this will be the normalisation of ${\bf n}\times \hat{{\bf v}}$ in our case (with $\hat{{\bf v}}$ a unit vector in the direction of ${\bf v}$), then 
\begin{equation}
\cos\zeta = \hat{\bf e}\cdot{\bf n}\times{\bf l},\quad \sin\zeta = \hat{\bf e}\cdot{\bf l}.
\end{equation}
Assuming ideal MHD and following steps in Appendix C of \cite{lyutikov2005polarization} we find 
\begin{align}
\hat{{\bf e}} = \frac{{\bf n}\times{\bf q}}{\sqrt{q^2-({\bf n}\cdot{\bf q})^2}}\quad {\bf q}=\hat{\bf B}+ {\bf n}\times({\bf v}\times\hat{{\bf B}}),\\
\nonumber\hat{{\bf B}} = \frac{1}{\sqrt{1-(\hat{{\bf B}}'\cdot{\bf v})^2}}\left[\hat{{\bf B}}' - \frac{\gamma}{1+\gamma}\hat{{\bf B}}'\cdot{\bf v}\right],
\end{align}
where $\hat{{\bf B}}$ is the unit vector field of ${\bf B}$.

Using these Faraday variables we can calculate the fractional polarisation $F$ as
\begin{equation}
F = \frac{\sqrt{Q^2 + U^2}}{I}.
\end{equation}
\subsection{Faraday rotation}
Following \cite{clausen2011signatures} the Faraday rotation $R$ is 
\begin{equation}
R  = C_3\int_{s_1}^{s_2}n_T'{\bf B}'\cdot{\bf n}'\rmd{s}.
\end{equation}
Where $n_T'$ is the (jet frame) thermal density of electrons. In this note we follow arguments given in \cite{clausen2011signatures} that the central core of the jet is responsible for synchrotron emission, but observational evidence indicates the Faraday screen responsible for the rotation is not co-spatial with this region. They propose a density for which the thermal electron density increases  radially towards the edge of the jet, if $\rho$ is the radial coordinate of the jet then the proposal is that
\begin{equation}
\label{thermalden}
n_t'(\rho) = (\rho/k_1)^6\exp(-(\rho/k_1)^2),
\end{equation}
where $\rho$ is the radial distance form the jet's central axis and $k_1$ is chosen to fit the normalization of our tube (its radius is $1$, which is not the case in \cite{clausen2011signatures}). This is partly based on the choice of field in \cite{clausen2011signatures}, a reverse pinch whose jet axis field  reverses direction at $\rho/k_1=1$.
\subsection{The fields}
In what follows we represent the field in cylindrical coordinates. In practice the field is then rotated to align with the axis of the jet.
\subsubsection{The reverse field pinch ${\bf B}_r$}\label{revpinchintro}
\begin{figure}
\centering
\includegraphics[width=8cm]{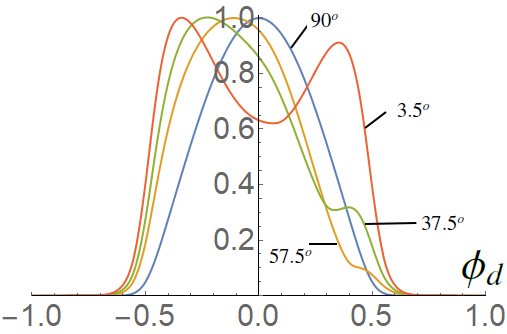}\quad\includegraphics[width=8cm]{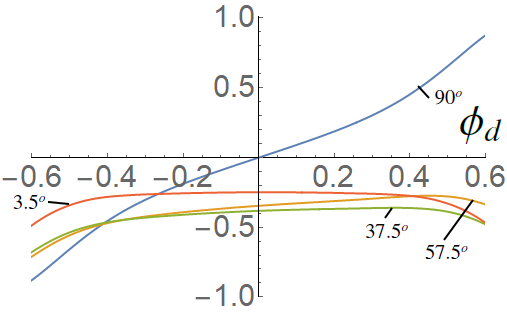}
\caption{\label{examples}Example $I$ (a) and $R$ (b) signatures for $\theta_d=0.5$ at jet angles $\theta_j=\pi/2,\pi/2-\pi/16,\pi/2-2\pi/16$ and $\pi/16$, which correspond to aberration corrected viewing angles of respectively  $90^o$,   $90^o,57.5^o,37.5^o,3.5^o$.}
\end{figure}
This is the field used in \cite{clausen2011signatures} and is used as a reference in this study. It can be written as
\begin{equation}
\label{revpinch}
{\bf B}_r= J_0(k_2\rho)\hat{z} + \left(x^2+y^2\right)^{-1/2}J_1(k_2\rho)\left(-y,x ,0\right)
\end{equation}
with $k_2=3.8312$, such that $J_1(k_2)=0$ and the field's rotation vanishes on the boundary. To match the Faraday cage  used in \cite{clausen2011signatures} (\textit{i.e.} \ref{thermalden}) we set $k_1=0.62761$. This field has a uniformly right handed rotational component, but the sign of the field's vertical component reverses at $\rho k_1=1$.

Calculations for the intensity $I$ and Faraday rotation $R$ for (\ref{revpinch})  for a slice of fixed $\theta$ values $\theta_d=0.5$ and $\phi_d\in[-1,1]$ are shown in Figure \ref{examples}. These are calculated $\gamma=10$ and jet angles $\theta$ which are similar to those used in \cite{clausen2011signatures}. They have the same basic shape as those of, respectively, Figures 2 and 4(b) of \cite{clausen2011signatures}, except that the plots are antisymmetric about the $y$-axis (their $\phi$ observational) behaviour is reversed. The reason for this is that the effect of viewing the jet from a significant distance in our model means we view it at an effective viewing angle in the form ${\bf n} = \left(-\sin(\xi),0, \cos(\xi)\right)$, where $\xi$ is small. This contrasts to ${\bf n} = \left(\sin(\xi),0, \cos(\xi)\right)$ (as would have been used in \cite{clausen2011signatures})  which would be more akin to an arm of a jet which is rotated  from a negative direction rather than a positive. So in our choice of the orientation of the jet's helical field, the line-of-sight component at $\phi_m$ is opposite to that of \cite{clausen2011signatures}. %
The significant asymmetry results from the fact that for a  helical geometry  tilting the jet towards the observer will mean on  one side  of the field will appear close to normal to the observer, whilst the other side will appear close to parallel. This leads to a relative boosting of the signals  either side of the central viewing direction. What is interesting here is the observation that, for a  given field chirality (handedness), this asymmetry depends which arm of the jet is pointing  toward the observer
as well as the field chirality. We stress that the handedness would not necessarily be obvious unless the jet's origin can be determined by observations, this is necessary to determine whether the photon propagation vector is {\bf ${\bf n} = \left(\sin(\xi),0, \cos(\xi)\right)$}, or ${\bf n} = \left(-\sin(\xi),0, \cos(\xi)\right)$. Because of this, the choice of the slice along which the observational quantities are measured needs to be perpendicular to the jet direction and the location of jet's origin needs to be taken into account. These can be easily determined for a radio galaxy jet seen side-on. However, in blazar observations appearing as circular intensity contours the jet's direction needs to be determined by combining further high resolution and time evolution observations \citep{gabuzda2018jets, Lister:2018}.

\subsubsection{The co-axial cable ${\bf B}_c$}
\begin{figure}
    \centering
    (a)\includegraphics[width=6cm]{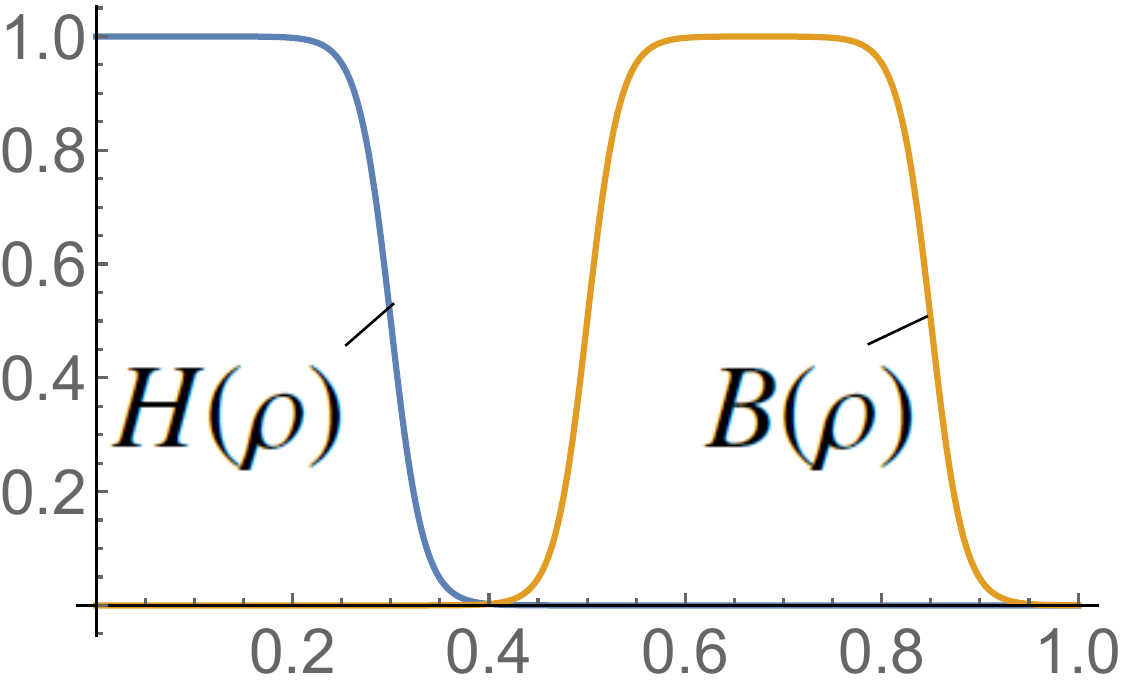}\quad(b)\includegraphics[width=6cm]{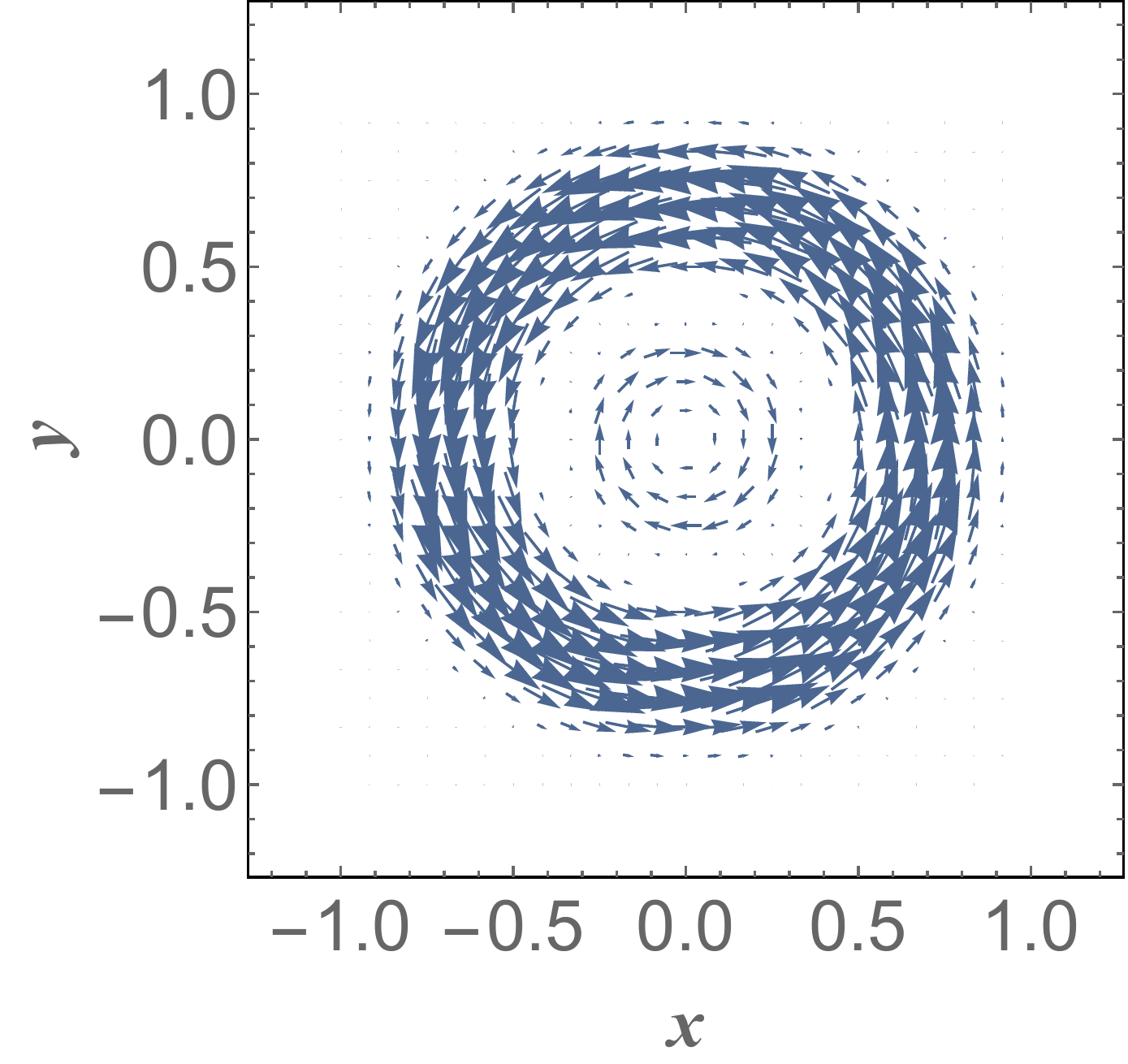}
    \caption{Representations of the co-axial cable field ${\bf B}_c$. (a) represents the functions $H(\rho)$ and $B(\rho)$ used to define spatially distinct co-axial twists. (b) an  in plane representation of the vector fields showing the two sheaths of opposing twist.}
    \label{coaxialfigures}
\end{figure}
This is a representation of the coaxial-cable model (alternatively the battery model), proposed in \cite{Contopoulos:2006,gabuzda2018jets} as a  field structure based on the observation of Faraday rotation profiles. In this note we represent it mathematically as 
\begin{align}
 {\bf B}_c &= (y,-x,0)H(\rho) + (-y,x,0)B(\rho) + \hat{z}(H(\rho)+B(\rho)),\\
\nonumber\rho &= x^2+ y^2,\quad H(\rho)= \frac{1}{1+\mathrm{e}^{2 k_3\rho-0.3}},\\ \nonumber B(\rho)&= \frac{1}{1+\mathrm{e}^{-(2 k_3\rho-0.5)}}-\frac{1}{1+\mathrm{e}^{2 -(k_3\rho-0.85)}}
\end{align}
If $k_3$ is significant (we use $k_3=30$  in this study) then the function $H$ has roughly the shape of a smoothed Heaviside function (see Figure \ref{coaxialfigures}(a)) whose value is $1$ at zero, and the function $B(\rho)$ creates a smooth bump whose peak value (between $0.6$ and $0.8$) is $1$. The first two components of ${\bf B_c}$ are then twisted fields with opposing chirality (see Figure \ref{coaxialfigures}(b)), they are chosen such that the current (approximately proportional to the curl of ${\bf B}$) is negative at the field's centre and positive near its edge. In this case it has a left-handed rotation on $\rho \in[0,0.4]$ and a right-handed rotation on $\rho \in[0.4,0.9]$. The third component, which controls the $z$ component of the field is negative on $\rho \in[0,0.4]$ and negative on $\rho \in[0.4,0.9]$. The (helical) handedness of the field is opposite in the core and the sheath in comparison to the reverse-field pinch where the poloidal field changes sign but the toroidal does not. 

\begin{figure}
\centering
    (a)\includegraphics[width=6cm]{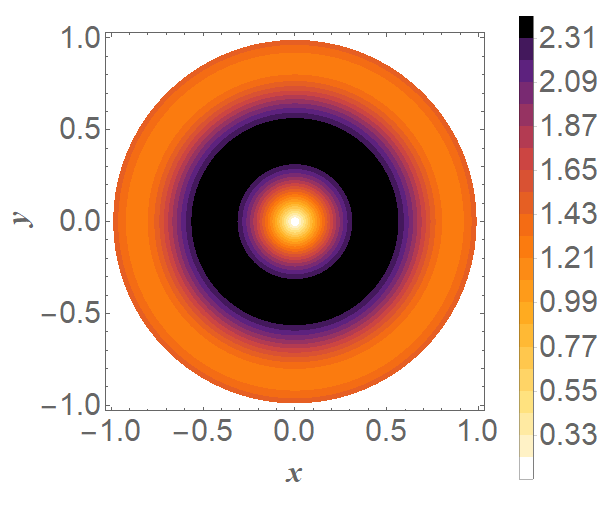}\quad(b)\includegraphics[width=6cm]{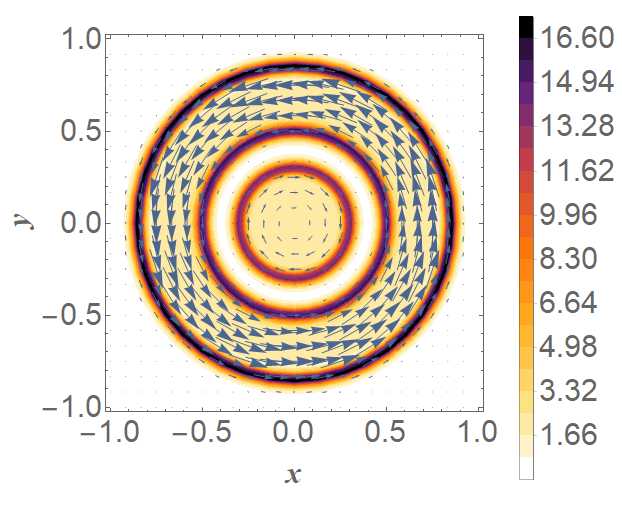}
    \caption{\label{twistcurrents} Contour plots of the magnitude of the curl $\nabla \times {\bf B}$ (a proxy for the current) in the field ${\bf B}_r$ (a) and the coaxial cable ${\bf B}_c$ (b). The field ${\bf B}_c$ is overlayed on (b). The maximum current in ${\bf B }_c$ is roughly an order of magnitude larger than in ${\bf B}_r$ which occurs in think layers at the edges of the two twisting domains.}
\end{figure}
A specific mathematical form for this coaxial cable/battery model is not given. The form we chose here, with non-overlapping and largely uniform twisting domains contrasts with that of the reverse pinch whose two regions of current blend smoothly into one another. One consequence of this would be the existence of significant thin sheets of current between the two twists. This is by comparison the the far more equally distributed current in the reverse pinch field, \textit{c.f.} (a) an (b) of Figure \ref{twistcurrents}. In principle we expect the coaxial cable field to obtain some dynamical state within the jet, which could give it a smoother profile, and we use the above mathematical expression as a framework to explore its effect on the observations. 

\subsubsection{A braided field {\bf B}}
\begin{figure}
(a)\includegraphics[width=6cm]{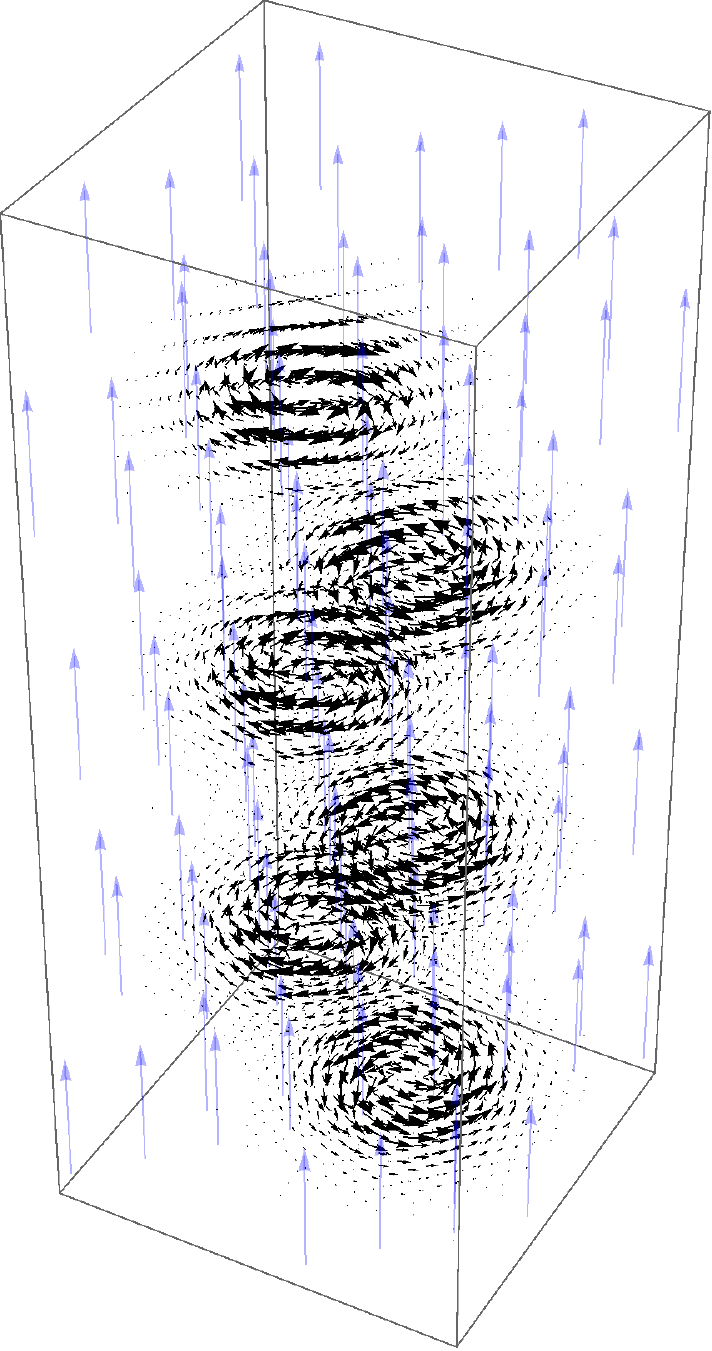}.
\caption{\label{braidfig}A visualisation of the braided field used in this study. Units of (exponentially decaying) twist are arranged to overlap. The vertical background field is also shown.}
\end{figure}
This field has been used as an initial condition in numerous simulations in a non-relativistic MHD context \cite{pontin2011dynamics,wilmot2010dynamics,prior2016twisted}. The field is composed of exponential twist units ${\boldsymbol B}_t(b_0,k,a,l,x_c,y_c,z_c)$ given by
\begin{align}
  \label{bt}
  \nonumber &{\bf B}_t(b_0,k,a,l,x_c,y_{c},z_{c})\\ &=\frac{2 b_0 k}{a J}\mathrm{exp}\left(-\frac{(x-x_{c})^2+(y-y_{c})^2}{a^2}-\frac{(z-z_c)^2}{l^2}\right){\bf R},\\
 \nonumber  {\boldsymbol R}&= (-(y-y_c),(x-x_c),0),
\end{align}
where the parameter $b_0$ determines the strength of the field, $a$ the horizontal width of the twist zones, $l$ their vertical extent and $k$ the handedness of the twist ($k=1$ is right handed). The centre of rotation is $(x_{c},y_{c},z_{c})$. The braided field is then defined as a superposition of $n$ pairs of positive and negative twists and a uniform vertical background field 
\begin{align}
  \label{braidfieldtube}
        {\bf B}_{b}(1,a,l,d,R,n) &= \sum_{i=1}^{n}{\bf B}_t(1,1,a,l,0,-d,s_d i)\\
\nonumber &+ {\bf B}_t(1,-1,a,l,0,d,s_d(i+1))  + b_0\hat{z},
\end{align}
where, $d$ is the offset from the jet's axis, and $s_d$ is the vertical spacing between consecutive twists (of  the same sign). An example is shown in Figure \ref{braidfig}. In this study the values we use are $a=\sqrt{2}/5,l=2/48,d_s=1/3,b_0=1,d=1/5$, these values are those used in \cite{pontin2011dynamics,wilmot2010dynamics,prior2016twisted} scaled proportionally to a domain of unit width. This field has significantly complex field line entanglement (caused by the staggered opposing twist structure) and its evolution leads to a diffuse current structure of small but significant current sheets. It can be shown to relax to a force-free state which is not a Taylor state \textit{e.g.} \cite{wilmot2010dynamics}. In a solar context it might be imagined to be produced either by a series of convective cells or entanglement due to complex mixing motion at its foot points. Here we simply use it as a field whose structure is significantly different form the twisted distributions, but whose complex structure is on a similar order of magnitude to the resolution of the observation (a fact we demonstrate in what follows).

\section{Results}
\label{Results}

Here, we explore the observational signatures varying the magnetic field structure, viewing angle and Lorentz factor. We have used two Lorentz factors $\gamma=10$ for the highly relativistic case and $\gamma=2$ for the mildly relativistic. We consider viewing angles between $\pi/2$ (edge on), $\pi/3$ (tilted) and $\pi/30$ (Blazar).
In the highly relativistic case aberration effects mean the tilted field is viewed at an angle closer to $28^{\rm o}$.
We perform our calculations assuming the ratio of the observer distance $d_0$, to the jet height is $2\times 10^5$, which is equivalent roughly to a jet of 1 arc second extent if seen side-on. We did experiment with varying this ratio, but unless it is unrealistically small there is little qualitative effect in comparison to the results reported here. Thus, our results though could be re-scaled for jets of arbitrary angular extent. We consider a jet of cylindrical shape and symmetry, where  its height to radius ratio is $10$.

We create synthetic images of both the Faraday rotation $R$ and the fractional polarization $F$, in both cases these are  over-layed on contours of the radio intensity $I$. 

Using the procedure detailed in section 2, we calculate the  distributions on a grid of $50\times 50$. To mimic the effect of beam convolution, we apply a Gaussian matrix of standard deviation $0.1$ along the x-direction and $0.2$ along the $y$ direction (which represents twice the scaled distance). These standard deviations are the same as used in \cite{clausen2011signatures}. We refer to these smoothed quantities as  $F_s,I_s$ and $R_s$ in what follows. Typically the beam size will be larger than this (\textit{e.g.}~\cite{asada2002helical, Hovatta:2012, Gabuzda:2015}), we ran calculations with larger beam sizes and found no difference from the results presented (save a rescaling of the pattern dimensions), so in some sense these results indicate the (current) best case scenario for radio observation.

For the sake of clarity we normalise the quantities $R,R_s, I$ and $I_s$ between $[-1,1]$, the dimensionless fractional polarization results are not normalised.

\subsection{Reverse pinch vs Coaxial cable}
\begin{figure}
(a)\includegraphics[width=7cm]{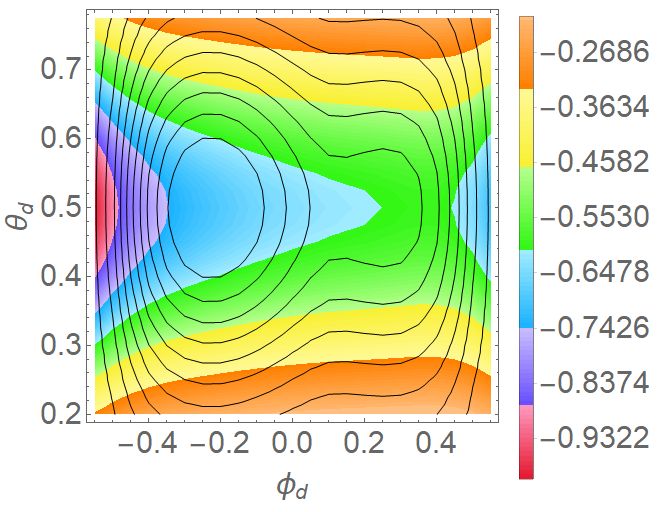}\quad (b)\includegraphics[width=7cm]{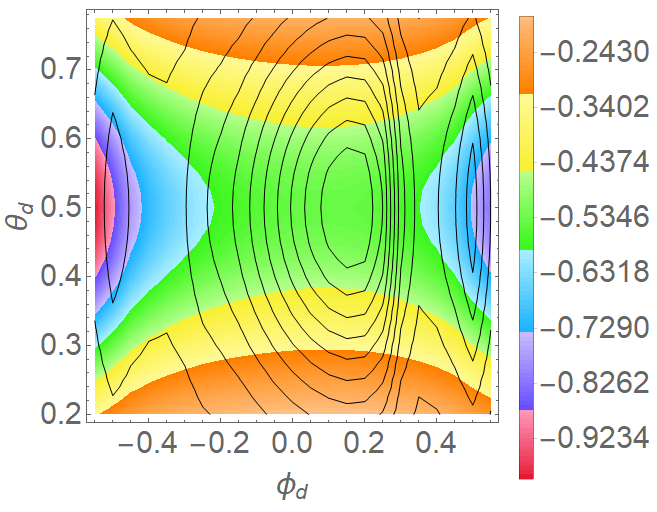}
\caption{\label{blazartwisthighrelFaradayblazar}. Contour plots of the intensity in black and the Faraday rotation profiles shown in colour of the two large scale helical fields for highly relativistic jets  $\gamma=10$ with viewing angle $\theta_j=\pi/3$. (a) $R_s$, for the field ${\bf B}_r$, (b)  $R_s$, for the field ${\bf B}_c$. The horizontal and vertical axes are scaled so that they they cover the extend of $\phi_d $ and $\theta_d$ as described in section \ref{BASIC_GEOMETRY}.}
\end{figure}
\begin{figure}
(a)\includegraphics[width=7cm]{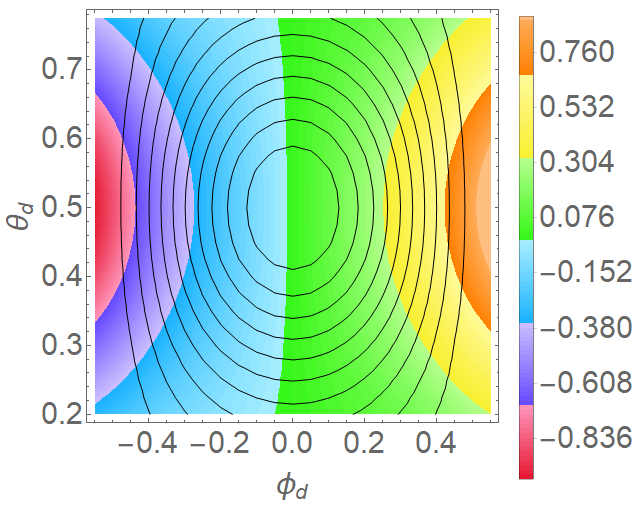}\quad
(b)\includegraphics[width=7cm]{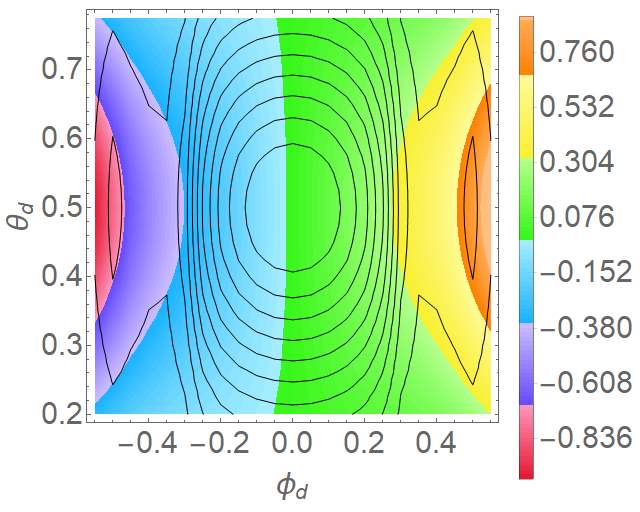}
\caption{\label{blazartwisthighrelFaradaysideon} Contour plots of the intensity in black and the Faraday rotation profiles shown in colour, of the two large scale helical fields for highly relativistic jets  $\gamma=10$ with viewing angle $\theta_j=\pi/2$. (a) $R_s$, for the field ${\bf B}_r$, (b)  $R_s$, for the field ${\bf B}_c$.}
\end{figure}
\begin{figure}
(a)\includegraphics[width=7cm]{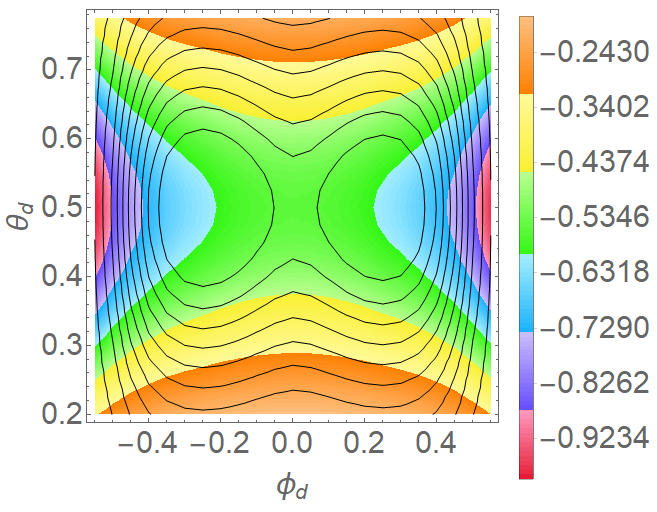}\quad
(b)\includegraphics[width=7cm]{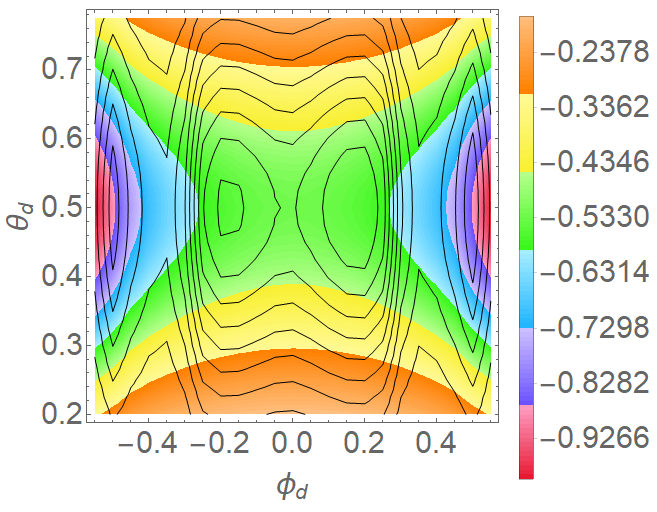}
\caption{\label{actblazartwisthighrelFaraday} Contour plots of the intensity in black and the Faraday rotation profiles shown in colour, of the two large scale helical fields for highly relativistic jets  $\gamma=10$ with viewing angle $\theta_j=\pi/30$. (a) $R_s$, for the field ${\bf B}_r$, (b)  $R_s$, for the field ${\bf B}_c$.}
\end{figure}
\begin{figure}
(a)\includegraphics[width=7cm]{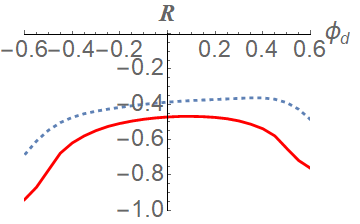}\quad
(b)\includegraphics[width=7cm]{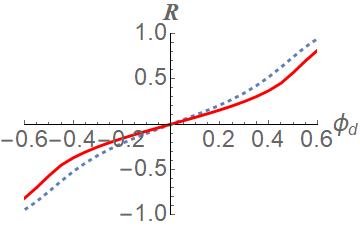}
\quad
(c)\includegraphics[width=7cm]{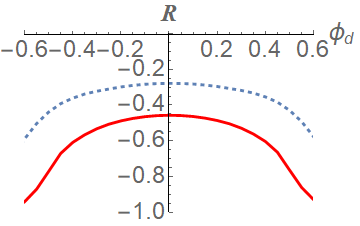}
\caption{\label{gradientstwist}Slices of the (smoothed) Faraday rotation $F_s$ at $\theta_d=0.5$ for the fields ${\bf B}_r$ (dashed) and ${\bf B}_c$ (solid). (a) the tilted case. (b) the side on case. (c) the Blazar case.}
\end{figure}

The Faraday rotation profiles of both ${\bf B}_r$ and ${\bf B}_c$ in the highly relativistic case $\gamma=10$ (the value used in  \cite{clausen2011signatures}) can be seen for viewing angle $\theta_j=\pi/3$ in Figure \ref{blazartwisthighrelFaradayblazar}(a). The ``side-on'' case $\theta_j=\pi/2$ (no relativistic aberration) is shown in  Figure \ref{blazartwisthighrelFaradaysideon}(a) . In both cases there is an asymmetry present in $R_s$, it is somewhat more pronounced in the side-on case. What is perhaps surprising is that the gradient of the curves is the same in both cases. Specific 1-D slices are taken at $\theta_d=0.5$ in both the tilted and side-on cases which illustrate this, these plots are shown in Figure \ref{gradientstwist}. The coaxial gradients are more pronounced, as one might expect due to the fact that its rotation does not decay smoothly like the reverse pinch field, rather it is mostly constant as a function of radius (except over a small range). The Blazar case $\theta_j=\pi/30$ is shown in Figure  \ref{actblazartwisthighrelFaraday}. In all three cases there is a minimal asymmetry along the $\phi_d$ direction a fact made clear in the 1-D slices are taken at $\theta_d=0.5$  (Figure \ref{gradientstwist} (c)). Of course if the slice were chosen for only part of this domain (say $\phi_d \in[0,0.6]$) then one \textbf{would} see a gradient.  This would also be true if the slice were taken at an angle rather than for fixed $\theta_d$.
  
\begin{figure}
(a)\includegraphics[width=7cm]{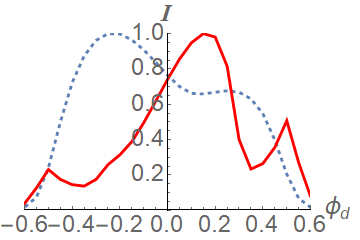}\quad
(b)\includegraphics[width=7cm]{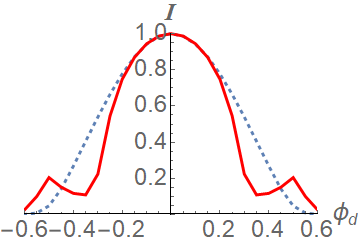}\quad 
(c)\includegraphics[width=7cm]{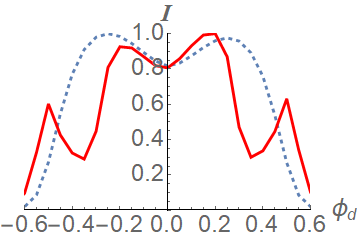}
\caption{\label{gradientstwistorig}Slices of the (smoothed) Synchrotron radiation Intensity $I_s$ at $\theta_d=0.5$ for the fields ${\bf B}_r$ (dashed) and ${\bf B}_c$ (solid). (a) $\theta_j=\pi/3$, (b) $\theta_j=\pi/2$ and (c) $\theta_j=\pi/30$.}
\end{figure}
As expected the contours of $I_s$ \textbf{do} show opposing asymmetry for the two fields in the tilted case (see \cite{clausen2011signatures}), see Figure \ref{gradientstwistorig}(a) as well as the less obvious the Blazar case ( Figure \ref{gradientstwistorig}(c)). The Blazar case has two peaks close to the centre of the observers viewpoint, this is also clear in Figure \ref{actblazartwisthighrelFaraday}. It is interesting to observe a number of Blazar observations have this twin peak contour structure, whereas some don't (see \textit{e.g.} \cite{gabuzda2018jets}). The relatively sharp transition in twisting of the coaxial field leads to a slightly more interesting sub-structure in the contours of $I_s$, as indicated in Figures \ref{gradientstwistorig}(a) and (c). The side-on case, as expected shows no asymmetry for either field (Figure \ref{gradientstwistorig}(b)). So we can distinguish the two different field chiralities if the field is tilted, but we have to interrogate the Intensity profiles in that case. Further, as discussed in section \ref{revpinchintro}, we could only make this distinction if we know the position of the jet origin with respect to the observer and the jet's tip. In the side on case it would appear from this information to be difficult to discriminate between globally twisted fields of two opposing chiralities.

\begin{figure}
(a)\includegraphics[width=7cm]{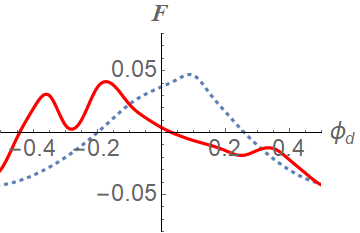}\quad
(b)\includegraphics[width=7cm]{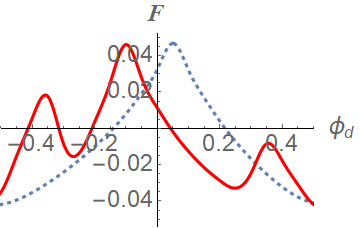}\quad 
(c)\includegraphics[width=7cm]{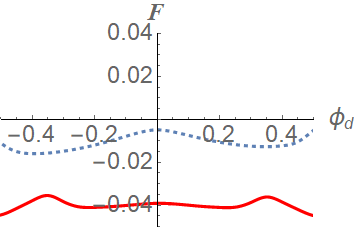}
\caption{\label{gradientstwistlp}Slices of the (smoothed) fractional polarization intensity $-F_s$ at $\theta_d=0.5$ for the fields ${\bf B}_r$ (dashed) and ${\bf B}_c$ (solid). (a) the tilted field,(b) the side on field, (c) the Blazar.}
\end{figure}

Slices of the fractional polarization $F_s$ are shown in Figure \ref{gradientstwistlp}(a) for the tilted jet ($\theta_j=\pi/3$)  the side-on jet (b) and the Blazar jet (c) (we show the negative of these values for easy comparison to the results of \cite{clausen2011signatures}).  The side on and tilted case have clear asymmetry and the two chirality fields have opposing gradients, {\bf the Blazar case is symmetric for both fields (we also tried an angle $\pi/16$ which \textbf{did} show some asymmetry)}. Unlike for the intensity $I_s$ the $F_s$ for side-on jets has the property of asymmetry. As with the intensity plots there is an additional structure in the coaxial cable where the exponential  twisting transitions occur.

\begin{figure}
(a)\includegraphics[width=7cm]{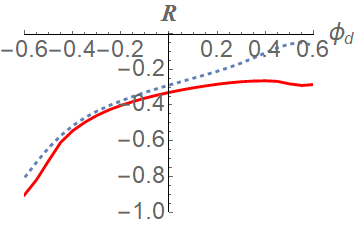}\quad (b)\includegraphics[width=7cm]{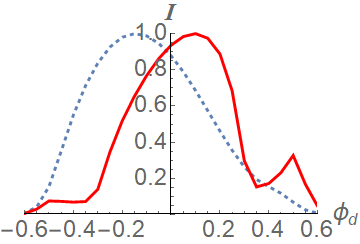}\quad (c)\includegraphics[width=7cm]{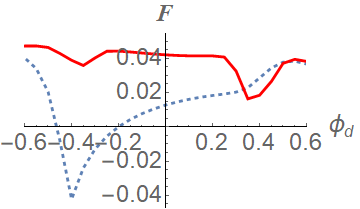}
\caption{\label{weakgammatwist}A comparison of slices of $R_s$ (a), $I_s$ (b) and $F_s$ (c) for the helical fields ${\bf B}_r$ (dashed) and ${\bf B}_r$ in a tilted jet $\theta_j=\pi/3$ with $\gamma=2$. The slices are taken at $\theta_d=0.5$. }
\end{figure}
\begin{figure}
(a)\includegraphics[width=7cm]{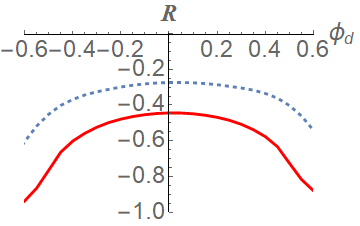}\quad (b)\includegraphics[width=7cm]{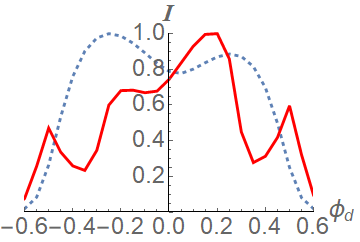}\quad (c)\includegraphics[width=7cm]{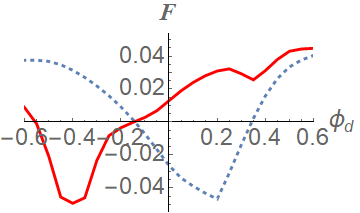}
\caption{\label{weakgammatwistBlazar}A comparison of slices of $R_s$ (a), $I_s$ (b) and $F_s$ (c) for the helical fields ${\bf B}_r$ (dashed) and ${\bf B}_r$ in a Blazar jet $\theta_j=\pi/30$ with $\gamma=2$. The slices are taken at $\theta_d=0.5$. }
\end{figure}

We now consider similar calculations in the weakly relativistic case $\gamma=2$. We restrict to the tilted and Blazar cases as they affected by the value of $\gamma$. We see in Figures \ref{weakgammatwist} and  \ref{weakgammatwistBlazar} the story is largely similar with the Faraday rotation profiles still appearing as similar (although the gradients are more distinct) and the intensity distinguishing the field's chirality. Finally the fractional polarisation $F_s$ distribution is asymmetric with opposing gradients for  the two fields (as in the highly relativistic case). {\bf The only difference from the highly relativistic case is that the fractional polarization slices \textbf{do} show asymmetry for the Blazar jet.}

\subsection{Braiding}
\begin{figure}
\resizebox{\hsize}{!}{\begin{tabular}{cc}
(a)\includegraphics[width=7cm]{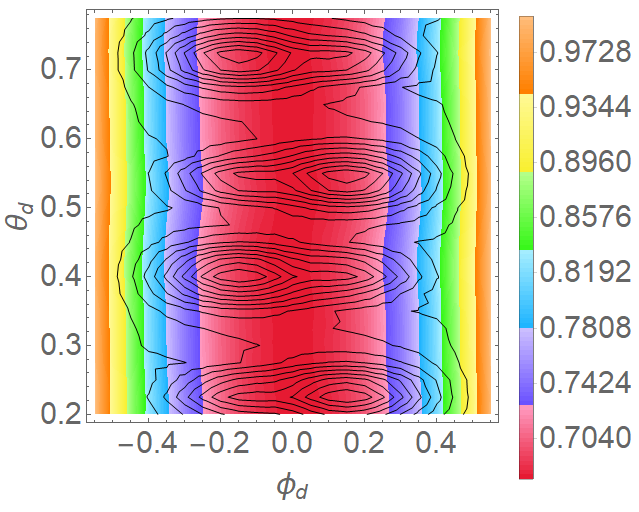} & (b)\includegraphics[width=7cm]{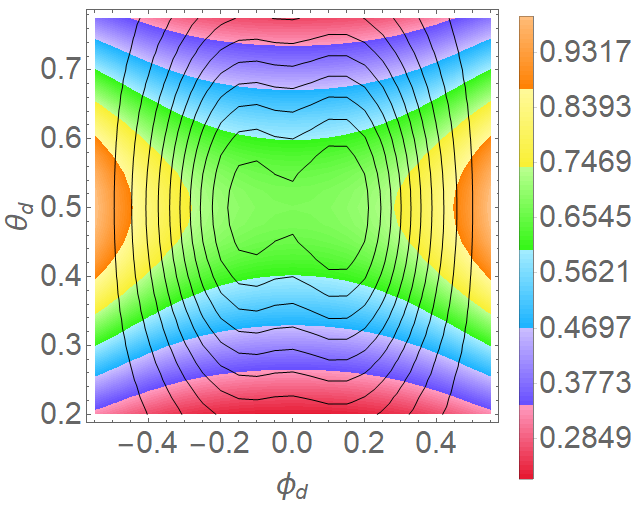}
\end{tabular}}
\caption{\label{blazarbraidhighrelFaraday} Faraday rotation profiles for the braided field $B_b$ at the tilted viewing angle $\theta_j=\pi/3$. (a) $R$, (b) $R_s$.}
\end{figure}
\begin{figure}
\resizebox{\hsize}{!}{\begin{tabular}{cc}
(a)\includegraphics[width=7cm]{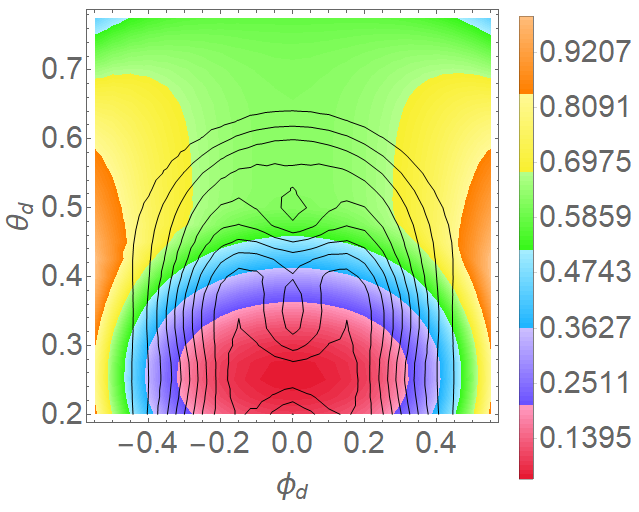} & (b)\includegraphics[width=7cm]{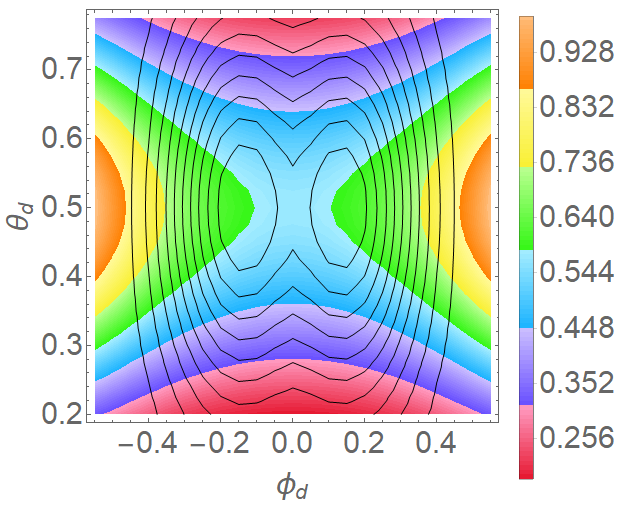}
\end{tabular}}
\caption{\label{actblazarbraidhighrelFaraday} Faraday rotation profiles for the braided field $B_b$ at the tilted viewing angle $\theta_j=\pi/30$. (a) $R$, (b) $R_s$.}
\end{figure}
\begin{figure}
(a)\includegraphics[width=7cm]{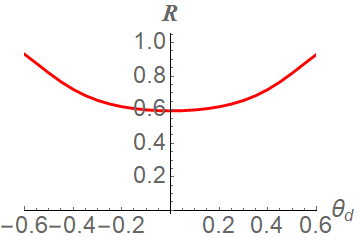}\quad (b)\includegraphics[width=7cm]{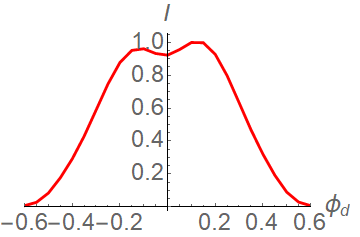}\quad 
(c)\includegraphics[width=7cm]{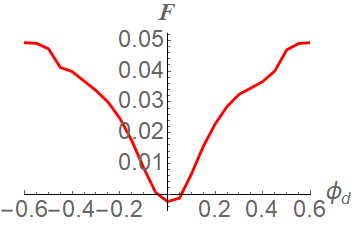}
\caption{\label{blazarbraidhighrelFaradayslice} Slices of the distributions $R_s$ (a) and  $I_s$ (b) and $F_s$ (c) at $\theta_d=0.5$ for the braided field $B_b$ at the tilted viewing angle.}
\end{figure}
\begin{figure}
(a)\includegraphics[width=7cm]{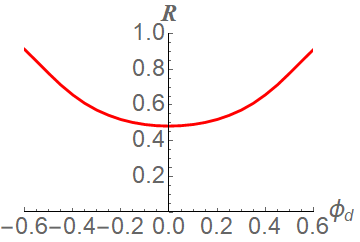}\quad (b)\includegraphics[width=7cm]{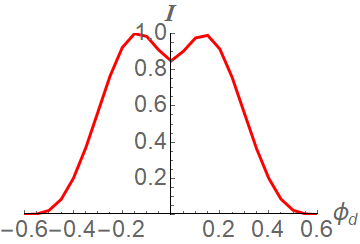}\quad 
(c)\includegraphics[width=7cm]{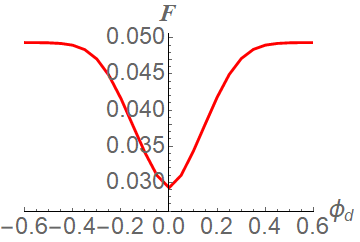}
\caption{\label{ActblazarbraidhighrelFaradayslice} Slices of the distributions $R_s$ (a) and  $I_s$ (b) and $F_s$ (c) at $\theta_d=0.5$ for the braided field $B_b$ at the Blazar viewing angle.}
\end{figure}
\begin{figure}
\resizebox{\hsize}{!}{\begin{tabular}{cc}
(a)\includegraphics[width=7cm]{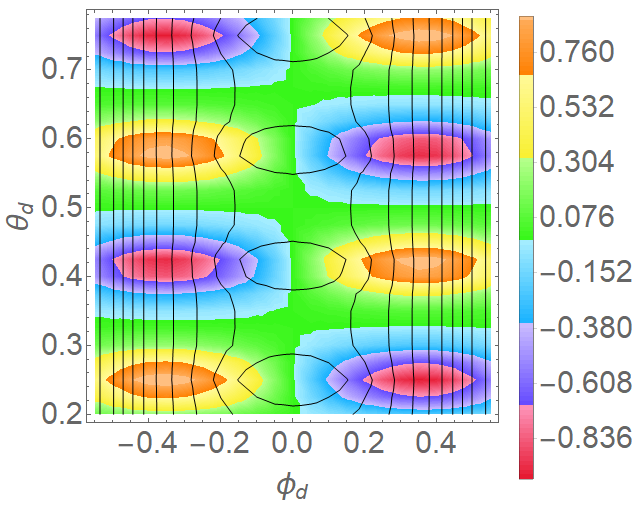} & (b)\includegraphics[width=7cm]{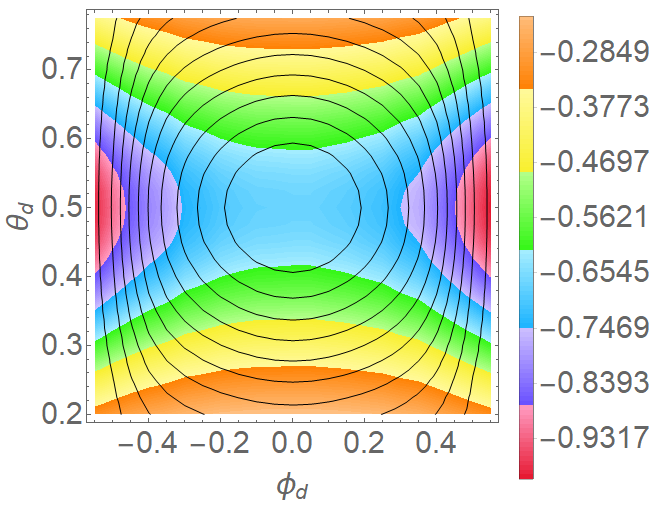}
\end{tabular}}
\caption{\label{sidebraidhighrelFaraday}  Faraday rotation profiles for the braided field $B_b$ at the side-on viewing angle $\theta_j=\pi/2$. (a) $R$, (b) $R_s$.}
\end{figure}

The Faraday rotation profiles of ${\bf B}_b$ in the highly relativistic case $\gamma=10$ can be seen for a viewing angle $\theta=\pi/3$ in Figure \ref{blazarbraidhighrelFaraday}. We also plot the pre-Gaussian smoothed distributions $I$ and $R$ to highlight the loss of information which occurs dues to the low  resolution of observations. The Faraday rotation profiles are (almost) symmetric (see Figure \ref{blazarbraidhighrelFaradayslice}(a)). A similar story is true of the Blazar case, Figure \ref{actblazarbraidhighrelFaraday} where the pre-smoothed Faraday rotation profiles show no asymmetry along the $\phi_d$ direction (although there is some sense of asymmetry along the $\theta_d$ direction, this is lost upon applying the Gaussian filter. We note in this case the Faraday profiles are remarkably similar to those of the reverse pinch (Bessel) field shown in Figure \ref{actblazartwisthighrelFaraday}(a).

In both the tilted and Blazar case the complexity of field is present in the pre-smoothed contours of $I$ (Figure \ref{blazarbraidhighrelFaraday}(a) and Figure \ref{actblazarbraidhighrelFaraday}(a)). Much of this information is lost upon applying the Gaussian filter and replaced with a set of slightly asymmetric contours in the tilted case Figure \ref{blazarbraidhighrelFaraday}(b), as highlighted in Figure \ref{blazarbraidhighrelFaradayslice}(b). In the Blazar case there is perfectly asymmetric bi-modal peaked distribution (see  Figures \ref{ActblazarbraidhighrelFaradayslice}(b) and \ref{ActblazarbraidhighrelFaradayslice}(b)) Similarly, as indicated in Figures \ref{blazarbraidhighrelFaradayslice}(c) and\ref{ActblazarbraidhighrelFaradayslice}(c) the fractional polarization (c) is essentially symmetric for both viewing angles (actually symmetric in the blazar case).  We emphasize that the Blazar Intensity distribution of this braided field is qualitatively the same as that of the reverse pinch field (\textit{c.f.} Figure \ref{ActblazarbraidhighrelFaradayslice}(b) and Figure \ref{gradientstwistorig}(c)), however, the fractional polarization distribution is symmetric where for the reverse pinch field is asymmetric (\textit{c.f.} Figure \ref{ActblazarbraidhighrelFaradayslice}(c) and Figure \ref{gradientstwistlp}(c)).

The same distributions are shown for the side-on case $\theta_j=\pi/2$ in Figure \ref{sidebraidhighrelFaraday}. Pre-smoothing (Figure \ref{sidebraidhighrelFaraday} (a)) the braided structure is clear in the Faraday rotation profiles (unlike in the tilted case shown in Figure \ref{blazarbraidhighrelFaradayslice}(a)), which actually show a series of alternating gradients with height. The intensity contours indicate the spatial variance of the field, but  the varying offset twist structure is lost due to line of sight averaging . However, almost all of this information is lost in the smoothing process as shown in Figure \ref{sidebraidhighrelFaraday}(b), which shows gradients in neither $I_s$ or $R_s$. We do not show the fractional polarization profile as it is qualitatively similar to the tilted case (symmetric). Further we do not report on the results  for $\gamma=2$ as they tell a qualitatively similar story.

The main conclusion we take here is that radio observations can hide significant complexity in  magnetic field structure as a consequence of both line-of sight averaging and the limited spatial resolution of observations. It is interesting to note that in the side on case it is the Gaussian convolution which removes the Faraday rotation structure, whilst, for the tilted and Blazar jet case much of that structure is already lost due to the line of sight averaging. We will revisit this issue in section \ref{losstype}.

\subsubsection{Asymmetry and braiding}
\begin{figure}
\resizebox{\hsize}{!}{\begin{tabular}{cc}
(a)\includegraphics[width=7cm]{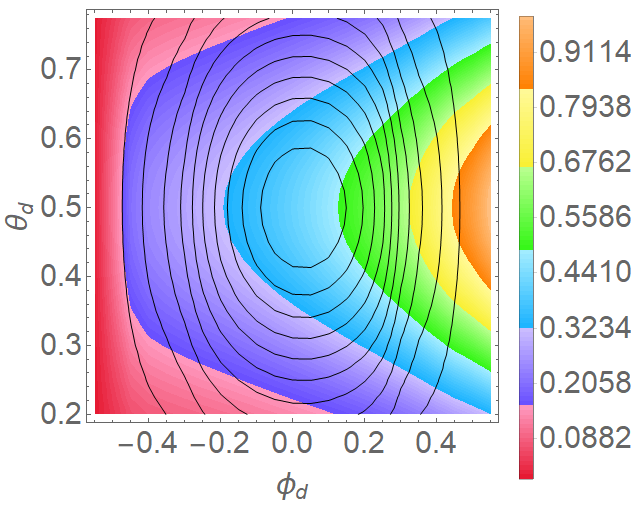} & (b)\includegraphics[width=7cm]{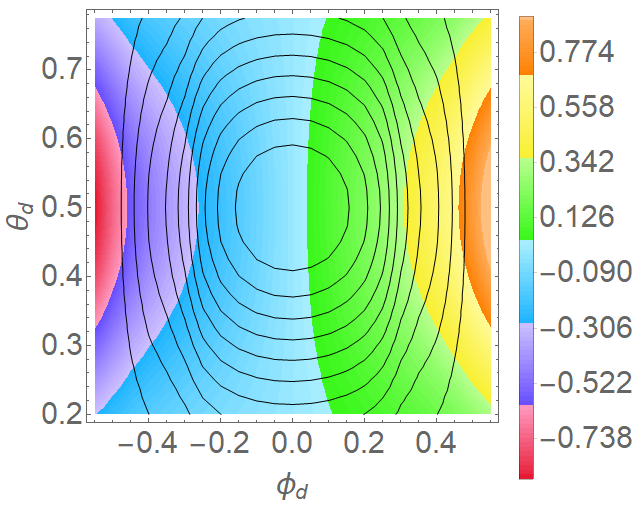}
\end{tabular}}
\caption{\label{twistandbraidcomp} Distributions of the (smoothed) Faraday rotation profiles $R_s$ for the field ${\bf B}_c+{\bf B}_b$, in the case (a) of a tilted jet with $\gamma=2$, (b) a side-on jet with $\gamma=10$.}
\end{figure}
\begin{figure}
\includegraphics[width=7cm]{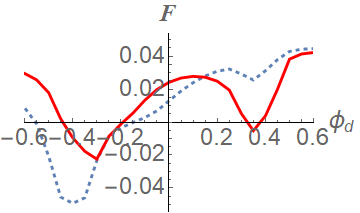}
\caption{\label{linpolcoaxialcompbrtw}Slices of the fractional polarisation at $\theta_d=0.5$ for the coaxial cable ${\bf B}_c$ (dashed) and the mixed field ${\bf B}_c+{\bf B}_b$, both for a weakly relativistic ($\gamma=2$) blazar field $\theta =\pi/30$}
\end{figure}

For non-Blazar type observations  Faraday rotation data is sufficient to distinguish the braided field from the large scale helical fields which show asymmetry in their profiles (although as discussed above, the Faraday rotation struggles to make the distinction between braided and straight fields). For blazar fields we can still discriminate between the two fields types, but only with the fractional polarization data {\bf (and even then this depends on the value of $\gamma$)}. One might then ask what would happen if the jet field had a mixture of the two field types? In Figure \ref{twistandbraidcomp} we see the Faraday rotation profiles $R_s$ for the composite field ${\bf B}_c+{\bf B}_b$ (note the fields are of a same order of magnitude at their maximum). This covers both side-on and blazar jets. The Faraday rotation profiles consistently shows the asymmetry of the twisted field. In (a) ($\gamma=2,\theta_j=\pi/3$) we see the asymmetry of profile of I. As for the potential for Blazar distinction, the fractional polarization distribution is qualitatively similar to the coaxial cable case (an example is shown in Figure \ref{linpolcoaxialcompbrtw}), in the sense they both have a similar asymmetry.

The crucial point here is that one could easily observe the kind of gradients expected of a large scale helical field but miss the signature of an additional component (of significant scale) which would lead to a far more complex jet magnetic field topology.

 \section{Types of structure loss}\label{losstype}
\begin{figure}
\resizebox{\hsize}{!}{\begin{tabular}{cc}
 (a)\includegraphics[width=7cm]{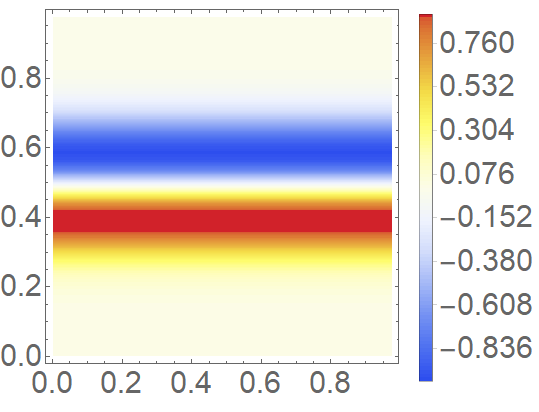} & (b) \includegraphics[width=7cm]{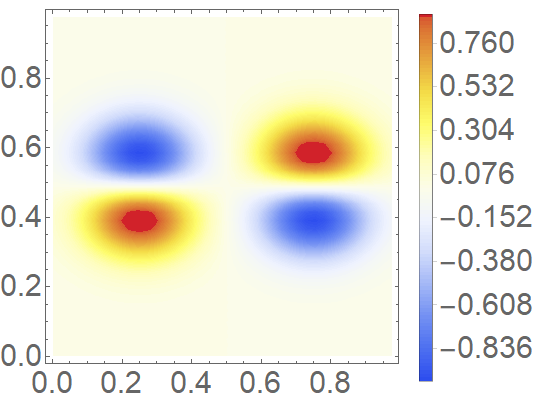}
\end{tabular}}
\caption{\label{missingpairs}Examples of the observationally lost distributions $I_0$. Panel (a) is a striped pattern, which could represent a gradient in the field. Panel (b) has a pattern similar to that given in the Faraday rotation profiles of the braided field ${\bf B}_b$}
\end{figure}
\begin{figure}
\centering
\includegraphics[width=6cm]{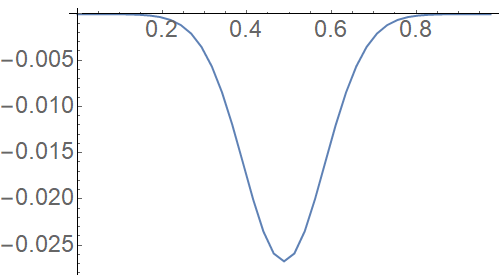}
\caption{\label{gaussianvec}A depiction of the ``geometry" vector ${\bf w}$.}
\end{figure}
\begin{figure}
\centering
\includegraphics[width=7cm]{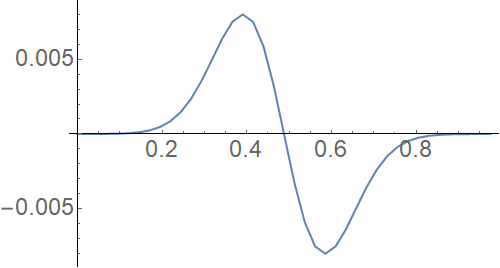}
\caption{\label{normdist1}A vector ${\bf w}_n$ which is normal to the vector ${\bf w}$ depicted in Figure \ref{gaussianvec}.}
\end{figure}
As discussed above there are several sources of information loss when synthetic observations are produced. One in particular takes  simple  form. The equation for converting a (discretely sampled) distribution $I,F$ or $R$, \textit{i.e.} a matrix (here we assume it is an $n$-by $n$ matrix) into the observable (matrix) distributions $I_s,L_s$ and $R_s$, takes the form (\textit{e.g.} for $I$)
\begin{equation}
I_s = G I,
\end{equation}
where $G$ is a Gaussian matrix.
Mathematically the simplification of $I_s$ occurs due to the singular nature of the matrix $G$. Thus we could write this equation in the form 
\begin{equation}
\label{mateq}
I_s = G(I_s^{-1} + I_0) = G I_s^{-1}.
\end{equation}
where the matrix $I_0$ is the part of he signal $I$ which is in the kernel of  $G$ ($GI_0 ={\bf 0})$, the information which is lost, and $I_s^{-1}$ the part which is required to produce the observed signal (it can can uniquely be determined by inversion (\textit{e.g.}  \cite{horn1990matrix}). In short matricies $I_0$ represent the distributions which are lost due to the finite beam width of the observational instrument. We have already seen in the previous section that braided patterns are, if not completely lost, significantly simplified by this transformation. Shortly we will show that the set of ``lost" patterns which belong to $I_0$ include striped and ``spotted"  such as shown in Figure \ref{missingpairs}. But first we discuss the potential implications of this fact.

\subsection{Balanced annulment and net zero-helicity fields} 
As indicated in Figure \ref{missingpairs} the distributions which are not observable tend to have an equal amount of positive and negative density . This occurs for the braided field ${\bf B}_b$ which has a balanced amount of both positive and negative twisting. It can be shown (\textit{e.g} \cite{russell2015evolution}) that this means its magnetic helicity, a quantity which measures the average entanglement of the magnetic field lines \citep{prior2014helicity}, is zero, even though it has a complex entanglement. In ideal or close to ideal Magnetohydrodynamics the magnetic helicity is conserved, hence the braided field through its evolution maintains an equal balance of twisting \citep{russell2015evolution}. Thus in an evolution of a (close to) ideal plasma, if complex structure is created (this could be on top of a net helicity field like a helical field), we should expect it to be balanced (in terms of {\bf its twisting structure}). The results of this paper indicate this would produce emissive signatures for which the net zero helicity field structure will either be annulled or at the least significantly masked.

\subsection{Characterising the lost information $I_0$ mathematically}
 We can get a useful idea of some of the potential structures of the observationally lost matrices $I_0$. The vector space of the matrix $I_0$ is very large as the Gaussian matrix only has one degree of freedom, along the diagonal (\textit{i.e.} the radial direction). A productive approach to describing this space is to use the singular decomposition form of  the matrix $G$ \citep{horn1990matrix}, which takes the form
\begin{equation}
G = UMV^*
\end{equation}
 where $M$ is a diagonal matrix (the singular equivalent of the eigenvalue matrix) and both $V$ and $U$ are unitary matrices. In this case all the matrices are real and $M$ has only one non-zero entry (the first), which represents the above mentioned radial degree of freedom. Thus the matrix $MV^*$ has non zero entries only in the first row. Let us call this row (vector) ${\bf w}$ and assume it is $n$-dimensional. A plot of the coefficients of ${\bf w}$ is shown in Figure \ref{gaussianvec}, (we have scaled the $x$-axis between $[0,1]$ since this vector acts on columns of $I_0$, which represent the $\theta_d$ direction). Note it takes the shape of a slice across a Gaussian, thus represented the above discussed radial degree of freedom. We can thus construct a matrix $I_0$ whose columns are $n$-dimensional and which are normal to ${\bf w}$, we label such vectors ${\bf w}_n$; they are drawn from an $n-1$ dimensional subspace. One way to make a simple basis for such vectors is to define vectors in the form 
\begin{equation}
{\bf w}_n^i = \left(0,0,\dots -w_{i+1},w_i,\dots 0\right)
\end{equation}
\textit{i.e.} for a given $i$ we swap the $i^{th}$ and $(i+1)^{th}$ entry of ${\bf w}$ and make the $i^{th}$ entry of the new vector  the negative of what is was, then make all other entries $0$. Thus we can write
\begin{equation}
{\bf w}_n  =\sum_{i=1}^{n-1}a_i{\bf w}_n^i. 
\end{equation}
for some set of real coefficients $\left\{a_i\right\}_{i=1}^{n-1}$.
For example if we chose all coefficients to be equal ($a_i=C$) then the annulling vector ${\bf w}_n$ has the geometry of the derivative of ${\bf  w}$ (see Figure \ref{normdist1}(a)). Assuming this is the same for each column of $I_0$ gives an ignored distribution in the form of, a vertically asymmetric gradient (as shown in Figure \ref{missingpairs}(b)). The symmetry of the Gaussian matrix means the same would be true of vertical stripes. But vertical stripes in this context would mean a gradient similar to that observed for the Faraday rotation (the decay outside the centre of the curve would result from decaying synchrotron emission outside of some jet core. The fact that in practice we don't see the loss of such gradients is because the helical structure is of a scale larger than the size of the observation beam. 

If we instead formed a matrix whose columns took the form shown in  Figure \ref{normdist1}(a), but whose coefficient values $C$ varied across the columns (the $\theta_d$ direction) like a pair of Gaussian's
\begin{equation}
\mathrm{e}^{-20(\theta_d-0.25)^2} -\mathrm{e}^{-20(\theta_d-0.75)^2} .
\end{equation}
then we obtain a lost distribution very similar to the type expected for both $I$ and $R$ of the braided field, as shown in
 Figure \ref{missingpairs}(b). Of course in practice we saw the braided field has some measurable effect on the smooth distributions, this is because they were not always quite on the scale required to be fully annuled. The point here is that distributions which are asymmetric with respect to either the central $\theta_d$ or $\phi_d$ axes of the jet can be either lost or severely diluted as long as they are of a similar order of magnitude to the beam size.    This might, for example make it hard to expect to detect a reverse gradient of the coaxial cable, as proposed by \cite{gabuzda2018jets}.

\section{Discussion}

Following the analysis in section \ref{Results} and 4 we compare the conclusions of our exploration of magnetic field profiles with relativistic jet observations. As we have studied generic and simplified models our comparison will be qualitative rather than quantitative.  
\begin{figure}
\includegraphics[width=8cm]{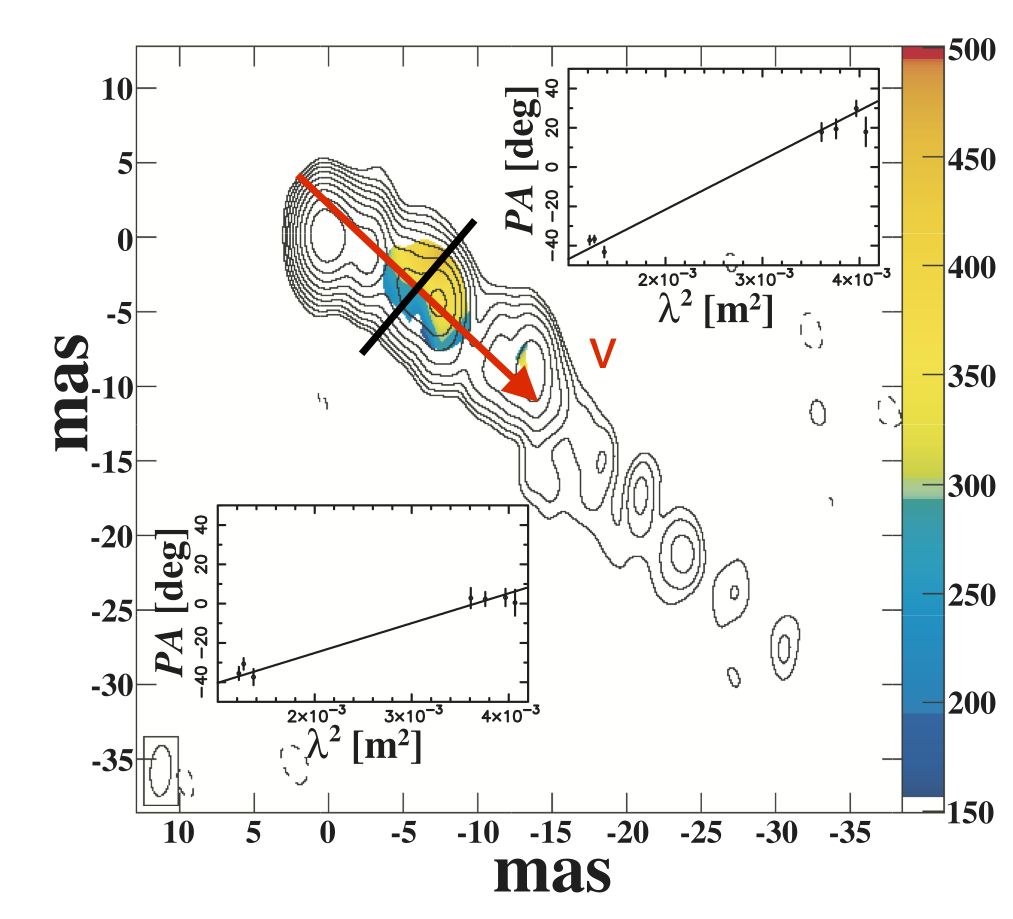} \includegraphics[width=8cm]{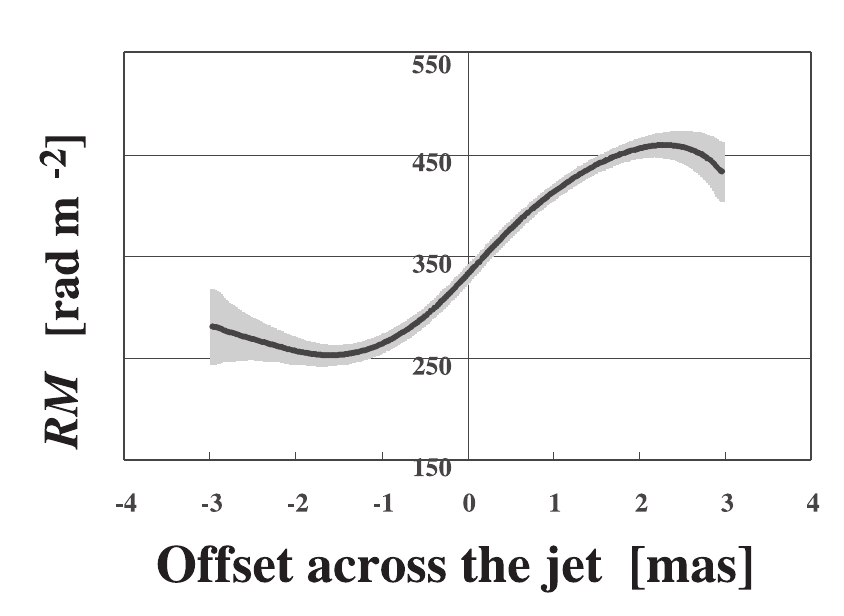}
\caption{\label{Asada} Top: Contours of radio emission of 3C 273 (black solid lines), with Faraday Rotation measure shown in colour. The jet propagation direction is shown with a red arrow. Bottom: Faraday Rotation measure along the black line, showing its transverse variation. Both figures reproduced by \citep{asada2002helical}. }
\end{figure}
\subsection{3C 273}

{\bf First, we consider 3C 273, a relativistic jet whose viewing angle is small $<16^{o}$ \citep{Abraham:1996,jorstad2017kinematics,liodakis2018constraining} and is believed to belong to the Blazar class}. Milliarcsecond radio observations demonstrate a clear variation of the Faraday rotation measure across the jet \citep{asada2002helical}, see Figures \ref{Asada}. In this system the jet propagation direction is easily identifiable, (shown with a red arrow), and the section along which the rotation measured is perpendicular to this direction. Based on the various models we have explored, one can see this type of of profile showing a gradient on a slice across the jet axis as in the helical or coaxial cable models (i.e. {\bf Figure}~\ref{blazartwisthighrelFaradaysideon}). Furthermore, this profile shows a counter-clockwise increase, thus if it is due to a coaxial cable configuration, one would see the outer part of the jet. We also note that this is consistent with a helical field with positive-handeness.   

\subsection{Blazar 0552$+$398}

\begin{figure}
\includegraphics[width=8cm]{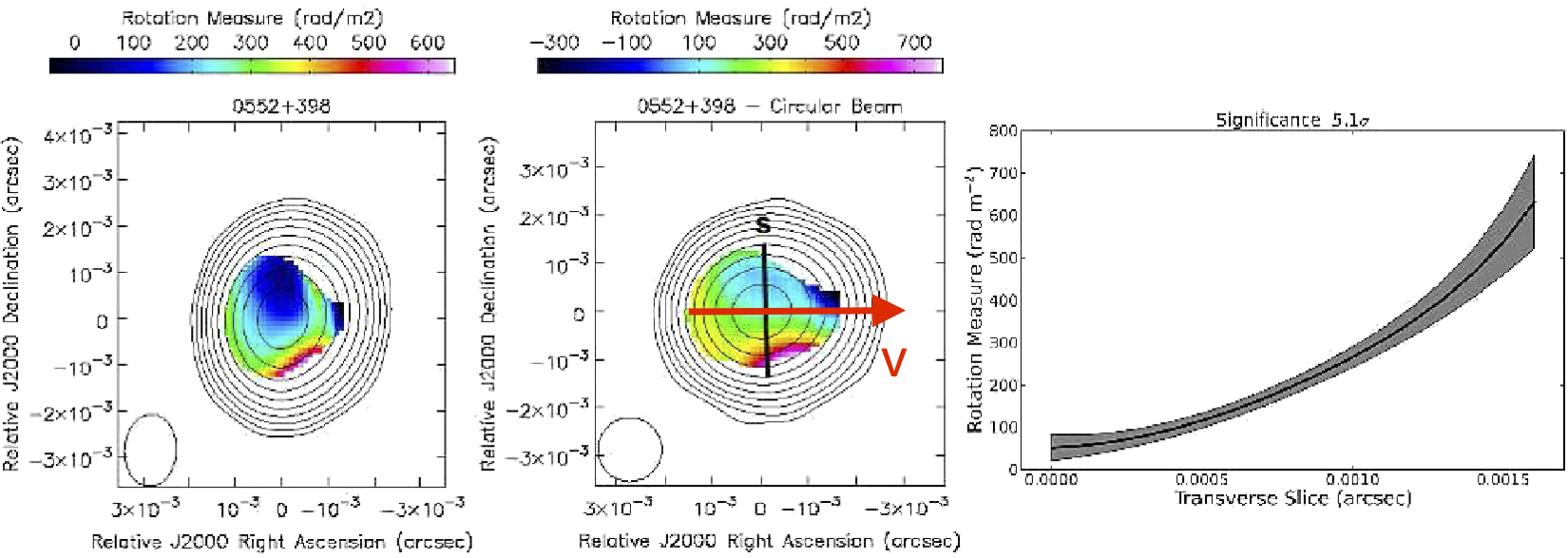} 
\caption{\label{Gabuzda} Radio emission contours (shown in black) and Faraday rotation measure of 0552$+$398, shown in color. The left panel is with an elliptical beam (shown at bottom left), the middle panel with a circular beam and the Faraday Rotation measure is shown in the right panel. The projected velocity on the plane of the sky is shown with a red arrow at the middle panel and the slice along which the Faraday Rotation is measures is plotted as a black line in the middle panel. Figure reproduced by \cite{gabuzda2018jets}. }
\end{figure}

Next, let us a consider the blazar source 0552$+$398 where the line of sight vector is almost parallel to the jet propagation direction \citep{gabuzda2018jets}. In this case the sky-projected jet propagation velocity is not as clearly identifiable as in 3C 273, however in combination with data from \cite{Lister:2018}, a propagation direction as shown in the red arrow of Figure \ref{Gabuzda} is favoured. In this case, one can attribute the rotation measure of this observation either to the inner part of a coaxial cable configuration, as its Faraday Rotation measure increases in the clockwise direction, or to a helical field with negative-handeness.  

We note here, that as pointed in \cite{clausen2011signatures}, the opening angle of blazar jets is comparable with the viewing angle, thus the simplified approximation of cylindrical jets reaches its limitations. 

\subsubsection{What could be missing?}
These conclusions should be tempered by the fact we have seen in the previous two sections that additional complex field structure would not necessarily be present in these observations. In particular we have seen that the addition of the braided field to a large scale helical field still yields the kind of gradients which are observed in these two cases. Although it is beyond the scope of this study, {\bf one would expect significantly different reconnective activity} if this more complex field were present, this could be manifested in high energy emission a possibility considered in \cite{blandford2017magnetoluminescence}.

\section{Conclusion}

From our exploration of jet magnetic field structure, Lorentz factor and viewing angle we can extract the following conclusions:

\begin{enumerate}
\item{It is possible that magnetic field structure whose spatial variation is at least of the order of magnitude to be hidden from the various optical observations of the jet, including the Synchrotron intensity, Faraday rotation and Linear polarization distributions. This is a consequence of the combination of the line of sight averaging and limited optical resolution.}
\item{If the position of the centre of the jet is not known as discussed in section \ref{revpinchintro}, then it  will not be possible to discriminate between large scale left and right handed helical fields.}
\item{Even if this relative orientation is known the Faraday rotation profiles do not discriminate the opposing chirality coaxial cable and reverse pinch large scale helical field structures.}
\item{If the field is almost side-on only the fractional polarization profiles (of the measures we consider here) can  discriminate the opposing chirality coaxial cable and reverse pinch large scale helical field structures.}
\item{Small islands of Synchrotron emission intensity contours could potentially indicate sharp changes in the magnetic field toroidal orientation.}
\end{enumerate}

Recent analysis of radio observations  \citep{Gabuzda:2015, gabuzda2018jets} have provided an ensemble of 52 sources where Faraday rotation measure can measured. Except for 5, the other 47 do not have any short time-scale variability and as of this, they can be used to identify a large scale magnetic field. According to that work 33 out of 47 have a rotation measure implying an electric current moving in the opposite direction of the jet propagation implying a coaxial cable. {\bf In our exploration of jet geometry we found that the side-on case shows clear gradient in the Faraday rotation }. On the other hand, the signatures of titled jets are rather sensitive to contribution from the edges of the jet. As this part of the jet also corresponds to lower intensity, there could be observational biases with the central part of the jet being clearly observable, but the edge, where the contribution of a toroidal field could be higher to be practically invisible. 

Finally, we remark that the gradient of the observable quantities could also depend on the choice of the slice. For instance a slice which is not normal to the jet's velocity could lead to a drastically different Faraday rotation jet profile. While the direction of the jet axis can be clearly identified in jets seen edge-on, a small deviation could have drastic implications in jets that are seen head-on and practically appear as concentric circles in observations.

\begin{acknowledgements}
    The authors would like to thank Daniele Dorigoni for a productive discussion regarding the singular value decomposition. We also thank Denise Gabuzda for permission to use Figure 3 from \cite{gabuzda2018jets} and discussion on the choice of the slice along which the RM is measured, and Maxim Lyutikov for insightful comments.
    \end{acknowledgements}



\bibliographystyle{aa} 
\bibliography{Bibtex.bib} 


\end{document}